\documentclass[structabstract]{aa}

\usepackage{graphicx}
\usepackage{natbib}
\usepackage{txfonts}
\usepackage[]{hyperref}
\usepackage{paralist}
\usepackage{multirow}
\usepackage{array}
\usepackage{subcaption}
\usepackage{arydshln}
\usepackage{xcolor}

\def\mb#1{{{#1}}}
\def\mbfth#1{{#1}}
\def\mbfth#1{{{#1}}}
\def\kv#1{#1}
\def\kvsec#1{#1}
\def\kvthrd#1{{{#1}}}
\def\kvfth#1{{{ #1}}}

\hypersetup{
    colorlinks,
    linkcolor=blue,
    citecolor=blue,
    urlcolor=blue,
}

\title{Hydrodynamical models of the $\beta$~Lyr~A circumstellar disc}

\author{
    K.~Vitovsk\'y\inst{\ref{prague}, \ref{SECOND}}       \and
    M.~Bro\v{z}\inst{\ref{prague}}
}
\institute{
     Charles University, Faculty of Mathematics and Physics, Institute of Astronomy, V~Hole{\v s}ovi{\v c}k{\'a}ch 2, 18000 Prague, Czech Republic
     \label{prague}
     \and
     \kvsec{Heidelberger Institut für Theoretische Studien, Schloss-Wolfsbrunnenweg 35, 69118 Heidelberg, Germany}
     \label{SECOND}
}

\date{Received x-x-2023 / Accepted x-x-2023}

\abstract
   {}
   {
We study dynamics of circumstellar discs,
with a focus on the $\beta$~Lyr\ae~A binary system.
This system with ongoing mass transfer has been extensively observed,
using photometry, spectroscopy and interferometry.
All these observations were recently interpreted using a radiation-transfer kinematic model.
   }
   {
We modified the analytical Shakura--Sunyaev models
for a general opacity prescription,
and derived radial profiles of various quantities.
\mb{These} profiles were computed for the fixed accretion rate,
$\dot M = 2\times 10^{-5}\,M_\odot\,{\rm yr}^{-1}$,
inferred from the observed rate of change of the binary period.
More general models were computed numerically,
using 1-dimensional radiative hydrodynamics,
accounting for viscous, radiative as well as irradiation terms.
The initial conditions were taken from the analytical models.
   }
   {
To achieve the accretion rate, the surface density~$\Sigma$ must be much higher
(of the order of $10^4\,{\rm kg}\,{\rm m}^{-2}$
for \mb{the viscosity parameter} $\alpha = 0.1$)
than in the kinematic model.
Viscous dissipation and radiative cooling in the optically thick regime
lead to a high midplane temperature~$T$
(up to $10^5\,{\rm K}$).
The accretion disc is still gas pressure dominated
\kv{with the opacity close to \kvthrd{Kramers} one.}
To reconcile temperature profiles \mb{with observations}, we had to distinguish three different temperatures:
midplane, atmospheric and irradiation.
The latter two are comparable to observations (30000 to 12000\,K).
We demonstrate that the aspect ratio~$H$ of 0.08 can be achieved in a hydrostatic equilibrium,
as opposed to previous works considering the disc to be vertically unstable.
   }
   {}


\begin{document}

\maketitle

\section{Introduction}

The $\beta$~Lyr~A system is an eclipsing binary undergoing rapid mass transfer due to the Roche-lobe overflow; the gainer in the system is an early B star of mass ${\approx}\,13\,M_\odot$ and the less massive donor a late B type star, ${\approx}\,2.9\,M_\odot$ \citep{2021A&A...645A..51B}.
The mass transfer has already switched their primary and secondary roles compared to the initial conditions \citep{1993A&A...279..131H}. 
This likely happened relatively recently, therefore the mass transfer rate is still close to its maximum \citep{2013A&A...557A..40D, 2024ApJ...966L...7L}.

The star was the second eclipsing binary discovered and the first ever Be star, with a more than 250 years long history of investigations; it is beyond the scope of this introduction to review it in detail (see \citealt{Harmanec2002}). 
Due to the lasting interest in this system, it serves as a prototype $\beta$-Lyr\ae-type binaries, that are considered natural laboratories of accretion, discs, mass transfer and mass loss.
They are one of the likely progenitors to stellar mergers and supernov\ae\ \citep{2024AJ....167...17C}.
Even recently this system is often mentioned as an example or comparison of possible dynamical phenomena in studies of other stars 
\citep{2023A&A...670A..94R,2023MNRAS.525.5121V,2022ApJ...936..129N,2022A&A...666A..51M}.
There is overwhelming evidence that binary mass transfer plays a key role also in the Be phenomenon for at least some systems \citep{2022MNRAS.516.3602E}, making $\beta$~Lyr a valuable target for dynamical modelling.

Evidence of a circumstellar material was first presented by \cite{baxandall1930spectrum}, who noticed additional redshifted absorption lines just before the middle of the primary eclipse.
The first direct interferometric observations were presented by \citet{2008ApJ...684L..95Z}.
Today, there is a broad consensus that the gainer is surrounded by an accretion disc \citep{10.1093/mnras/stt515,2018A&A...618A.112M,2021A&A...645A..51B}.
Observations show the binary's orbital period $P = 12.9440\,\textrm{days}$ (in 2020) \citep{2021A&A...645A..51B} is increasing at a rate of $\dot{P} = 19\,\textrm{s per yr}$, which corresponds to a mass transfer rate of $\dot{M} = 2 \cdot 10^{-5}\,M_{\odot}\,\textrm{yr}^{-1}$ \citep{Harmanec_1993A&A...279..131H}.
The binary also exhibits a second, much longer photometric cycle with a period of 275 days \kvthrd{\citep{1989SSRv...50...35G},} which \kvthrd{could be} caused by \kvthrd{optical depth} changes in the circumstellar disc \kvthrd{(\mb{a similar variability is seen in} AU Mon; \citet{2022A&A...664A.103A})}.
Actually, $\beta$~Lyr~A is atypical among the so-called double-periodic variable\kvthrd{s \citep{2017SerAJ.194....1M}} in that it has an evolving period value, while other such systems have a characteristically constant period \kvthrd{\citep{2023A&A...670A..94R}}.

The preferred distance is 328\,pc \citep{2021A&A...645A..51B}, which is slightly different from the associated cluster (294\,pc; \citealt{Bastian_2019A&A...630L...8B}).

The optically thick and thin circumstellar material in the system was recently studied by \cite{2018A&A...618A.112M} and \cite{2021A&A...645A..51B}, where a kinematic model was fitted to spectroscopic, light curve, spectral-energy-distribution (SED), interferometric and also differential-interferometric data.
While the accretion disc with a radius of ${\approx}\,31.5\,R_\odot$ is optically thick, it is surrounded by optically thin material, which results in variable emission-line profiles.
Already \citet{1996A&A...312..879H} proposed that some matter must be propelled in the direction perpendicular to the disc in the form of jets
, likely originating where the matter transported from the secondary interacts with the disc. The presence of the jets has been further developed by a number of authors \citep{2007A&A...463..233A,2013MNRAS.432..799M} and recently considered in the models of \citet{2021A&A...645A..51B}.

Hereinafter, we extend their work by adding equilibrium dynamics into consideration. We present an insight into the hydrodynamical state of the disc in the current phase of rapid mass transfer; even though we do not consider a long-term variability of $\beta$-Lyr\ae-type stars \citep{Mennickent_2021A&A...653A..89M,Rojas_2021ApJ...922...30R}.

In Sect.~\ref{sec:obs_param}, we describe the constraints we place on all our models.
In \kvthrd{Sect.~\ref{sec:anal_models}}, we give an overview of the derivation of a modification to the well-known analytical models of accretion discs by \citet{1973A&A....24..337S}.
The classical models generate radial profiles of hydrodynamical quantities of accretion discs around black holes in three different cases. 
One model assumes a dominance of the radiation pressure over the gas pressure and the Thomson scattering as the opacity source. 
The other two models assume dominance of the gas pressure; one again for an opacity arising from Thomson scattering and the other for \kvthrd{Kramers} opacity. 
We introduce a general opacity prescription and modify the set of equations so that our model better corresponds to a star as a central object.
\kvthrd{In Sect.~\ref{sec:num_mod} we introduce the \mb{numerical,} time-dependent radiation hydrodynamical \mb{model based on} \citet{2017A&A...606A.114C}.}
\kvthrd{In Sect.~\ref{sec:method} we describe how the \mb{models} presented in this work are \mb{scrutinised}.}
In Sect~\ref{sec: results} we present the \mb{specific results obtained for the $\beta$~Lyr~A disc.} 
Finally, in Sect.~\ref{discussion}, we \mb{discuss these} results \mb{in the context of other models and compare them to observations} presented by \citet{2021A&A...645A..51B}.

\section{Observational constraints of the $\beta$~Lyr~A disc}\label{sec:obs_param}

To constrain dynamical models of the $\beta$~Lyr~A disc,
we used profiles from recent studies of this stellar system,
conducted by \citet{2018A&A...618A.112M} \kvsec{and} \citet{2021A&A...645A..51B}.

In these studies, authors fitted a series of kinematic models
to an extensive set of observations.
Hereinafter, we use parameters and profiles from their sect.~3.3 ('joint compromise model') of \citet{2021A&A...645A..51B}, 
\kvthrd{the profiles may be constructed by inserting parameters from their Table 1. into equations 11.-15..}
\kvthrd{We refer to their results} simply as 'observations'.
Let us comment on some specific aspects of \citet{2021A&A...645A..51B}.

The temperature profile $T(r)$ is best constrained in the outer part of the disc (the system is observed edge-on).
The outer boundary temperature reaches just under 12\,000\,K,
the inner boundary temperature is close to the temperature of the star
(inferred from spectroscopy to be $T_{\star} \approx 30\,000\,{\rm K}$), 
where the slope of the profile is $-0.73$.
The surface density profile $\Sigma(r)$ ranges from approximately $1350\,{\rm kg}\,{\rm m}^{-2}$ at the inner boundary
to approximately $650\,{\rm kg}\,{\rm m}^{-2}$ at the outer boundary, with a slope of $-0.57$.
However, it should be interpreted as a lower limit,
because the total mass of an optically-thick medium are not well constrained
by observations of a surface (a.k.a. atmosphere).

During a detailed review of the model used in \citet{2021A&A...645A..51B},
we found that they assumed the mean molecular weight $\mu = 2.36$
and the adiabatic index $\gamma = 1.0$,
which are values typical rather for cool protoplanetary discs \citep{2016MNRAS.461.2257K}.
However, the circumstellar environment of the $\beta$~Lyr~A system is much hotter
and a composition closer to ionised hydrogen should be assumed.

Actually, one of the converged parameters was the multiplicative factor $h_{\rm cnb}$, defined as
\begin{equation}
    H = h_{\rm cnb} H_{\rm eq}\,,
\end{equation}
where $H$ denotes the actual pressure scale height of the disc 
and $ H_{\rm eq}$ the pressure scale height in a vertical hydrostatic equilibrium.
The best-fit value was $h_{\rm cnb} = 3.8$
and they interpreted this as a disc being vertically unstable.
We suspect that at least a part of the increase
is due to the choice of $\mu$ and $\gamma$.

Hence, we adjusted the values in \citet{2021A&A...645A..51B} model as follows,
$h_{\rm cnb} = 1.0$, $\mu = 0.5$ and $\gamma = 1.4$ \kvsec{as corresponds to ionised hydrogen and vertical hydrostatic equilibrium},
and recomputed the radial profiles of pressure scale height, density and pressure (from the converged temperature profile)\mb{;}
note that we did not run the convergence again.

The profiles are somewhat shifted, therefore we suspect the original high $h_{\rm cnb}$ has probably also other causes than the choice of $\mu$ and $\gamma$.  
We refer to these shifted profiles throughout the text as 'adjusted observations'.
Let us emphasise that we do not claim that the \citet{2021A&A...645A..51B} disc must be in vertical equilibrium, we use this modified model as a hydrostatic limit to the \citet{2021A&A...645A..51B} disc and we plot it for reference next to our results.

In our work, we adopted several values from \citet{2021A&A...645A..51B}
as fixed parameters (see Tab.~\ref{tab:fix_param}).
In addition, we take dynamics into consideration, the main dynamical constraint is the mass transfer rate between the two stars.
We assumed that the total mass is conserved during the transfer.

\begin{table}
\centering
\caption{Fixed parameters adopted from \citet{2021A&A...645A..51B}.}
\begin{tabular}{lcc}
\hline \hline Parameter & Unit & Value \\
\hline
$R_{\star}$ & $R_{\odot}$ & $5.987$ \\
$T_{\star}$ & $\mathrm{K}$ & 30000 \\
$M_{\star}$ & $M_{\odot}$ & $13.048$ \\
$q$ & 1 & $0.223$ \\
$\Dot{M}$ & $M_{\odot}\,\textrm{yr}^{-1} $ & $2 \cdot 10^{-5}$ \\
\hdashline
$R_{\text{innb}}$ & $R_{\odot}$ & $8.7$ \\
$R_{\text{outnb}}$ & $R_{\odot}$ & $31.5$ \\
\hline
\end{tabular}
\label{tab:fix_param}
\tablefoot{
$R_{\star}$~denotes the radius of the primary (gainer),
$T_{\star}$~its~effective temperature,
$M_{\star}$~its~mass,
$q$~the~mass ratio of the binary,
$R_{\text{innb}}$ the~inner boundary of the disc, and
$R_{\text{outnb}}$ the outer boundary, respectively. 
The parameters above the dashed line are fixed both in \citet{2021A&A...645A..51B} and here, the disc boundaries below the dashed line are converged results of the fit fixed in this work.
}
\end{table}

\section{Analytical models} \label{sec:anal_models}
In this section, we derive a modification to the analytical models of \citet{1973A&A....24..337S} to be better applicable to a case where the central object is a star (not a black hole).
\footnote{The Python scripts (Jupyter Notebooks) used to compute the models in this section are publicly available at \url{https://github.com/KristianVitovsky/Modified-Shakura-Sunyaev-models}.}

\subsection{Derivation of the analytical models}
Let us discuss the derivation in detail by listing all the equations, we followed the main steps of the derivation from \citet{1973A&A....24..337S} but for a more general case, adopting adjustments to the set of equations.
For a comprehensive description of assumptions behind the models see the original paper \citep{1973A&A....24..337S}.
The surface density $\Sigma$ is obtained by integrating the volume density over the vertical extent of the disc,
which is equal to multiplying by a factor of 2 times the disc's pressure scale height $H$,
\begin{equation}
    \Sigma = 2H\rho \, .
\end{equation}
The pressure scale height $H$ assuming hydrostatic equilibrium is given by the ratio of sound speed $c_s$ and the Keplerian angular velocity  $\Omega_{\rm K}$ \citep{Pringle_1981ARA&A..19..137P},
\begin{equation}
    H = c_{\rm s} r \sqrt{\frac{r}{GM_{\star}}} \, .
\end{equation}
For the sound speed a general formula in an isothermal approximation is \citep{1973A&A....24..337S}
\begin{equation}
    c_{\rm s} = \sqrt{\frac{P}{\rho}}  \, .
\end{equation}
We assume Keplerian rotation in the disc, then assuming that the central star will rotate with a sub-Keplerian velocity (otherwise it would be at its stability limit, but Be stars are shown to rotate at sub-critical velocities \citep{2013A&ARv..21...69R}) implies that the angular velocity curve  $\Omega(r)$ must have a maximum at or beyond the inner edge of the disc, that is $\frac{\partial \Omega}{\partial r}\bigg|_{r < R_{\rm innb}}  = 0$. This boundary condition then introduces a "bending" term into the equation for the surface density $\Sigma$ \citep{Pringle_1981ARA&A..19..137P},
\begin{equation} \label{eq:Sigma and bending}
     \nu \Sigma = \frac{\Dot{M}}{3 \pi} \left(1-\sqrt{\frac{R_{\star}}{r}} \right) \, ,
\end{equation}
where $\nu$ is the kinematic viscosity. 

Plugging into a more general equation for viscous energy dissipation $Q_{\rm visc}$, \citet{Pringle_1981ARA&A..19..137P} derives
\begin{equation}
    Q_{\text{visc}} = \frac{3GM_{\star}\Dot{M}}{8\pi r^3} \left(1-\sqrt{\frac{R_{\star}}{r}} \right) \, .
\end{equation}
The models assume radiative energy losses only from the two faces of the disc. This is given by the Stefan-Boltzmann law,
\begin{equation} \label{eq:q_{vert}}
        Q_{\text{rad}} = \frac{2\sigma_B T^4}{\tau_{\text{eff}}} \, ,
    \end{equation}
where the vertical temperature profile is approximated by a model atmosphere according to \citet{1990ApJ...351..632H} in an optically thick limit. The "effective" optical depth is then given by 
\begin{equation}
        \tau_{\text{eff}} = \frac{3}{8}\tau_{\text{opt}} ,
\end{equation}
the optical depth $\tau_{\rm opt}$ of the vertical layer is approximated as in \citet{2017A&A...606A.114C},
\begin{equation} \label{eq:tau_opt}
    \tau_{\text{opt}} = \frac{C_k}{2}\kappa \Sigma ,
\end{equation}
where $\kappa$ is opacity and the $C_k$ factor accounts for the opacity change above the midplane as suggested by comparisons of 3D and 2D models \citep{2012A&A...539A..18M}. In the case of our results, generated by these analytical models we always set the factor to $C_k = 0.6$, the same as is suggested for hydrodynamic simulations by \citet{2017A&A...606A.114C}. 

The models arise from an assumed fundamental energy balance, that is that all the energy generated by viscous friction is radiated away, this is given by 
\begin{equation} \label{eq:anal_ener_balance}
         Q_{\text{visc}} =  Q_{\text{rad}} \, .
\end{equation}

The kinematic viscosity $\nu$ is parameterised with $\alpha$, which can be introduced by assuming that angular momentum is exchanged between rings of the disc via eddies with a limit to size given by the vertical extent of the disc and velocity given by the local sound speed $c_s$ in the disc \citep{Pringle_1981ARA&A..19..137P},
\begin{equation} \label{eq:kin_visc_alpha_param}
    \nu = \alpha c_s H \, ,
\end{equation}
where $H$ is the disc pressure scale height.

As a generalisation to the models by \citet{1973A&A....24..337S} we used a more general form of an opacity prescription parameterised by three constants $\kappa_0$, $A$ and $B$,
\begin{equation} \label{eq:op_approx_general}
        \kappa = \kappa_0\,\rho^A T^B \, ,
\end{equation}
a choice of the constants then defines a specific model with a specific dominant opacity source.

The equation of state is the same as the one for stellar matter since the source of the matter is a binary mass transfer. It combines ideal gas and radiation pressure $P = P_{\rm g} + P_{\rm r}$ and is in its full form given by
\begin{equation} \label{eq: eq of state}
        P = \rho\frac{k_B T}{\mu m_p} + \frac{4\sigma_B}{3c} T^4 \, ,
\end{equation}
where $\sigma_B$ is the Stefan-Boltzmann constant, $c$ is the speed of light, $k_B$ the Boltzmann constant, $\mu$ the mean molecular weight, $m_p$ the mass of the proton, $T$ the midplane temperature and $\rho$ the midplane density. For simplicity, we assumed the disc is made of pure ionised hydrogen and chose $\mu = 0.5$. 

In order to obtain prescriptions for the radial profiles, we \kvthrd{can} apply different assumptions to the equation of state about the relative importance of gas pressure and radiation pressure. That is, we actually derive three different classes of models, each based on a single assumption, specifically $P_{\rm g} \approx P_{\rm r}$, $P_{\rm g} \gg P_{\rm r}$ and $P_{\rm g} \ll P_{\rm r}$ \kvthrd{.}

A careful inspection of the results of \citet{2021A&A...645A..51B} led us to consider all three assumptions as possibly \kvthrd{applicable} to the system. \kvthrd{\citet{2021A&A...645A..51B} profiles \mb{were} given for $\rho$ and $T$ (as referenced in Sec.~\ref{sec:obs_param}). Applying the equation of state (Eq.~(\ref{eq: eq of state})) gives contributions from gas and radiation pressure. The resulting pressure contributions are of approximately equal magnitudes.} However \kvthrd{we must further take into account}, the observed $\rho$ profile is a lower limit, hence the possibility of gas-pressure-dominated disc cannot be excluded. Moreover, there are still uncertainties in the $T$ profile\kvthrd{.} Especially in the inner part of the disc \citep{2018A&A...618A.112M} \kvthrd{the profile is weakly constrained}\kvthrd{,} due to the $\propto T^4$ proportionality, we should not a priori rule out radiation-pressure-dominated discs.

\subsection{Gas pressure dominated discs ($P_{\rm g} \gg P_{\rm r}$)} \label{gas_pressure_dom_model}

The first class is based on the assumption that the gas pressure dominates over the radiation pressure,
$P_{g \rm} \gg P_{r \rm}$.
Neglecting radiation pressure in Eq.~(\ref{eq: eq of state}),
we effectively recover the ideal gas equation of state.

To simplify notation, we denote the often-appearing sum of exponents as
\begin{equation}
    D = 3A - 2B + 10\, .
\end{equation}
Then,
\begin{gather} 
\kappa = \kappa_{\star}\alpha^{-\frac{7A+2B}{D}}\Dot{M}^{\frac{4(A+B)}{D}}M_{\star}^{\frac{(11A+6B)}{2D}}r^{-\frac{3(11A+6B)}{2D}} \left(1-\sqrt{\frac{R_{\star}}{r}} \right)^{\frac{4(A+B)}{D}}\, \label{mod_shaksun_gas_start} , \\
\kappa_{\star} = \left( \kappa_0 \rho_0^A T_0^B \right)^{\frac{10}{D}}\, , \\   
H = H_{\star}\alpha^{-\frac{A+1}{D}}\Dot{M}^{\frac{A+2}{D}}M_{\star}^{-\frac{A-2B+7}{2D}}r^{\frac{3(A-2B+7)}{2D}} \left(1-\sqrt{\frac{R_{\star}}{r}} \right)^{\frac{A+2}{D}}\, , \\
H_{\star} = H_0 \kappa_{\star}^{\frac{1}{10}}\, , \\
\Sigma = \Sigma_{\star}\alpha^{-\frac{A-2B+8}{D}}\Dot{M}^{\frac{A-2B+6}{D}}M_{\star}^{-\frac{A+2B-4}{2D}}r^{\frac{3(A+2B-4)}{2D}} \left(1-\sqrt{\frac{R_{\star}}{r}} \right)^{\frac{A-2B+6}{D}}\, , \\
\Sigma_{\star} = \Sigma_0 \kappa_{\star}^{-\frac{1}{5}}\, , \\
T  = T_{\star}\alpha^{-\frac{2(A+1)}{D}}\Dot{M}^{\frac{2A+4}{D}}M_{\star}^{\frac{2A+3}{D}}r^{-\frac{3(2A+3)}{D}} \left(1-\sqrt{\frac{R_{\star}}{r}} \right)^{\frac{2A+4}{D}}\, , \\
T_{\star}  = T_0 \kappa_{\star}^{\frac{1}{5}}\, \label{mod_shaksun_gas_end} ,
\end{gather}
where opacity-independent constants (subscript 0) are defined as
\begin{gather}
H_0 = \left( \frac{1.8}{256 \cdot \sigma_B \pi^2 G^{\frac{7}{2}}}  \left( \frac{k_B}{\mu m_p}\right)^4 \right)^{\frac{1}{10}} \, , \\
\Sigma_0 = \left( \frac{1.8}{256 \cdot \sigma_B \pi^2 G^{\frac{7}{2}}} \left( \frac{k_B}{\mu m_p}\right)^4 \right)^{-\frac{1}{5}} \cdot (\sqrt{G} 3\pi)^{-1}\, , \\
\rho_0 = \left( \frac{1.8}{256 \cdot \sigma_B \pi^2 G^{\frac{7}{2}}} \left( \frac{k_B}{\mu m_p}\right)^4 \right)^{-\frac{3}{10}} \cdot (\sqrt{G} 6\pi)^{-1}\, , \\
T_0 = \left( \frac{1.8}{256 \cdot \sigma_B \pi^2 G^{\frac{7}{2}}} \left( \frac{k_B}{\mu m_p}\right)^4 \right)^{\frac{1}{5}} \cdot \frac{\mu m_p G}{k_B}\, .
\end{gather}

We obtained prescriptions for the profiles of hydrodynamical quantities dependent on six input parameters. The form of the prescriptions is similar to the classical models. Each has four main parts; a scaling constant (with the index~$_{\star}$), scaling by some powers of the input parameters ($\alpha$, $M_{\star}$, $\Dot{M}$), a power law of the radial coordinate and the bending term introduced Eq.~(\ref{eq:Sigma and bending}).
For the sake of intuition, note that the bending factor always has the same exponent as the mass transfer rate $\Dot{M}$.
The exponents in each term are dependent on the parameters defining the specific opacity prescription.

The scaling constants don't have a specific meaning, yet for each separately, we may notice 
\kvsec{that far} away from the star, all prescriptions reduce to simple power laws of the radial coordinate.

\subsection{Gas and radiation pressures of equal magnitudes ($P_{g \rm} \approx P_{r \rm}$)}

The second class considers both sources of pressure of approximately equal importance,
$P_{g \rm} \approx P_{r \rm}$. In practice, we use 
\begin{equation}
    P = 2P_{g \rm} = 2P_{r \rm} \, .
\end{equation}

Again, we denote the often-appearing sum of exponents $D$, in this section, it corresponds to 
\begin{equation}
    D = 3A - 2B + 1\, .
\end{equation}
Then,
\begin{gather}
    \kappa = \kappa_{\star}\alpha^{-\frac{A}{D}}\Dot{M}^{\frac{2(B-A)}{D}}M_{\star}^{\frac{(2B-A)}{2D}}r^{\frac{3(A-2B)}{2D}} \left(1-\sqrt{\frac{R_{\star}}{r}} \right)^{\frac{2(B-A)}{D}}\, \label{mod_shak_sun_approx_start} , \\
    \kappa_{\star} = \left( \kappa_0 \rho_0^A T_0^B \right)^{\frac{1}{D}} \, , \\   
    H = H_{\star}\alpha^{-\frac{A}{D}}\Dot{M}^{\frac{A+1}{D}}M_{\star}^{\frac{(2B-A)}{2D}}r^{\frac{3(A-2B)}{2D}} \left(1-\sqrt{\frac{R_{\star}}{r}} \right)^{\frac{A+1}{D}}\, , \\
    H_{\star} = H_0 \kappa_{\star} \, , \\
    \Sigma = \Sigma_{\star}\alpha^{\frac{-A+2B-1}{D}}\Dot{M}^{\frac{A-2B-1}{D}}M_{\star}^{-\frac{A+2B+1}{2D}}r^{\frac{3(A+2B+1)}{2D}} \left(1-\sqrt{\frac{R_{\star}}{r}} \right)^{\frac{A-2B-1}{D}}\, , \\
    \Sigma_{\star} = \Sigma_0 \kappa_{\star}^{-2}\, , \\
    T = T_{\star}\alpha^{-\frac{2A}{D}}\Dot{M}^{\frac{2(A+1)}{D}}M_{\star}^{\frac{(2A+1)}{D}}r^{\frac{-3(2A+1)}{D}} \left(1-\sqrt{\frac{R_{\star}}{r}} \right)^{\frac{2(A+1)}{D}}\, , \\
    T_{\star} = T_0 \kappa_{\star}^{2}\, \label{mod_shak_sun_approx_end}  ,
\end{gather}
where opacity-independent constants are defined as
\begin{gather}
    H_0 = \frac{1.8}{16 \pi c}\, , \\
    \Sigma_0 = \frac{16^2 \pi c^2}{9.72 \sqrt{G}}\, , \\
    \rho_0 = \frac{16^3 \pi^2 c^3}{34.992 \sqrt{G}}\, , \\
    T_0 = \left( \frac{G\mu m_p}{2k_B} \right) \left( \frac{1.8}{16 \pi c} \right)^2\, .
\end{gather}
The obtained prescriptions are composed of the same terms as was described in Sec.~\ref{gas_pressure_dom_model}, yet the power law exponents contain other combinations of the opacity specific constants ($A$,$B$), the definitions of the scaling constants are also different.

\subsection{Radiation pressure dominated discs ($P_{\rm g} \ll P_{ \rm r}$)}

The third class are models of radiation-pressure dominated discs ($P_{g \rm} \ll P_{r \rm}$). As before we neglect the other part of the equation of state (Eq.~(\ref{eq: eq of state})), reducing it to the radiation pressure equation of state. Following \citet{1973A&A....24..337S}, to obtain a temperature profile we calculate the energy density $\epsilon$ of radiation inside a layer,
    \begin{equation}
        \epsilon = \frac{9}{32 \pi}\Dot{M}\frac{GM_{\star}}{r^3}\frac{\kappa \Sigma}{c}  \left(1-\sqrt{\frac{R_{\star}}{r}} \right) \, ,
    \end{equation}
    and plug in the Stefan-Boltzmann law,
    \begin{equation}
        \epsilon = \frac{4 \sigma_B}{c}T^4 \, .
    \end{equation}

Again, we denote the often-appearing sum of exponents $D$, in this section, it corresponds to
\begin{equation}
    D = 12A + B + 4\, .
\end{equation}
Then,
\begin{gather}
   \kappa = \kappa_{\star}\alpha^{-\frac{4A+B}{D}}\Dot{M}^{-\frac{8A}{D}}M_{\star}^{\frac{B-4A}{2D}}r^{\frac{3(4A-B)}{2D}} \left(1-\sqrt{\frac{R_{\star}}{r}} \right)^{-\frac{8A}{D}}\, \label{mod_shaksun_rad_start} , \\
   \kappa_{\star} = \left( \kappa_0 \rho_0^A T_0^B \right)^{\frac{4}{D}}\, , \\
   H_{\star}\alpha^{-\frac{4A+B}{D}}\Dot{M}^{\frac{4A+B+4}{D}}M_{\star}^{\frac{(B-4A)}{2D}}r^{\frac{3(4A-B)}{2D}} \left(1-\sqrt{\frac{R_{\star}}{r}} \right)^{\frac{4A+B+4}{D}}\, , \\
   H_{\star} = H_0 \kappa_{\star} \, , \\
   \Sigma = \Sigma_{\star}\alpha^{-\frac{4A-B+4}{D}}\Dot{M}^{\frac{4A-B-4}{D}}M_{\star}^{-\frac{4A+3B+4}{2D}}r^{\frac{3(4A+3B+4)}{2D}} \left(1-\sqrt{\frac{R_{\star}}{r}} \right)^{\frac{4A-B-4}{D}}\, , \\
   \Sigma_{\star} = \Sigma_0 \kappa_{\star}^{-2}\, , \\
   T = T_{\star}\alpha^{\frac{2A-1}{D}}\Dot{M}^{\frac{2A}{D}}M_{\star}^{\frac{4A+1}{2D}}r^{-\frac{3(4A+1)}{2D}} \left(1-\sqrt{\frac{R_{\star}}{r}} \right)^{\frac{2A}{D}}\, , \\
   T_{\star} = T_0 \kappa_{\star}^{-\frac{1}{4}}\, \label{mod_shaksun_rad_end} ,
\end{gather}
where opacity-independent constants for this class are defined as
\begin{gather}
    H_0 = \frac{1.8}{32 \pi c}\, , \\
    \Sigma_0 = \frac{32^2 \pi c^2}{9.72 \sqrt{G}}\, , \\
    \rho_0 = \frac{32^3 \pi^2 c^3}{34.992 \sqrt{G}}\, , \\
    T_0 = \frac{72 \sqrt{G} c^2}{9.72 \sigma_B}\, .
\end{gather}

Again, we obtained the same terms as in Sec.~\ref{gas_pressure_dom_model}, yet with different power-law exponents and different scaling constants.

\section{Numerical model} \label{sec:num_mod}

\kvthrd{In order to constrain the timescales of evolution, 
we also computed 1D hydrodynamical simulations}
of the $\beta$~Lyr~A disc
with an Eulerian solver on a polar staggered mesh
\citep{VANLEER1977276,1992ApJS...80..753S,2000A&AS..141..165M,2017A&A...606A.114C}.
\kvfth{\mbfth{Since our} analytical models are \mbfth{based} on a few strict assumptions, we use the numerical model to \mbfth{verify} their validity in this \mbfth{specific} case, \mbfth{as} described in Sec~\ref{sec:method}.}
We omit a comprehensive description of the code (FARGO, or FARGO\_THORIN),
but we introduce \kvfth{the respective set of fluid equations in Appendix~\ref{sec:hydroeq}.
Let us now note on some of the specifics of the model that are relevant for interpreting our results.}

\paragraph{Energy equation:}
The calculation of energy balance is more detailed in the numerical model
(cf.~Eq.~(\ref{eq:anal_ener_balance})).
Energy dissipation by viscous forces is determined by
\begin{equation}
Q_{\mathrm{visc}}=\frac{1}{2 v \Sigma}\left(\pi_{r r}^2+2 \pi_{r \theta}^2+\pi_{\theta \theta}^2\right)+\frac{2 v \Sigma}{9}(\nabla \cdot \boldsymbol{v})^2 \,,
\end{equation} 
where the viscous stress tensor $\boldsymbol{\pi}$ is calculated according to \citet{2002A&A...387..605M}.
Radiative losses $Q_{\rm rad}$ from the faces of the disc are described
also for optically thin matter, using the effective opacity
\citep{1990ApJ...351..632H,2017A&A...606A.114C},
\begin{equation} \label{eq:tau_with_thin}
    \tau_{\mathrm{eff}}=\frac{3}{8} \tau_{\mathrm{opt}}+\frac{1}{2}+\frac{1}{4 \tau_{\mathrm{opt}}} \, .
\end{equation}
This is then used in  the same Stefan–Boltzmann as in the analytical models (Eq.~(\ref{eq:q_{vert}})). Additionally, irradiation from the central star is included
and assumed to occur through the disc's faces \citep{2017A&A...606A.114C},
\begin{equation} \label{eq: Qirr}
Q_{\text{irr}}=\frac{2 \sigma_{\mathrm{R}} T_{\mathrm{irr}}^4}{\tau_{\mathrm{eff}}} \, ,
\end{equation}
where $T_{\rm irr}$ denotes the effective temperature of the radiation falling onto the disc's surface.
It is calculated by the projection of the radiative flux on the surface,
\begin{equation} \label{eq:Tirr}
T_{\mathrm{irr}}^4=(1-A)\left(\frac{R_{\star}}{r}\right)^2 T_{\star}^4 \sin \delta \, ,
\end{equation}
where $A$ denotes the disc's albedo and $\delta$ is an estimate of the incident angle \citep{2017A&A...606A.114C}.

\section{Methodology} \label{sec:method}

\kvthrd{To our aim, we explore the \mb{parameter} space of the modified Shakura-Sunyaev models (derived in Sec.~\ref{sec:anal_models}), in particular, the part that could be relevant to the studied $\beta$~Lyr~A \kvthrd{disc.} The results from the application of these analytical models are then used as initial conditions for simulations with the numerical \mb{model,} introduced in Sec~\ref{sec:num_mod}. The resulting radial profiles \mb{are} then compared to \citet{2021A&A...645A..51B}, both qualitatively and quantitatively. In this section we define the specifics of our approach.}

\subsection{\kvthrd{The} \mb{choice} of opacity} \label{subsec: choiceofop}

\kvthrd{When applying the analytical models to a system,} after choosing an assumption about pressure, the values of 6 input parameters ($\alpha$, $\Dot{M}$, $M_{\star}$, $\kappa_0$, $A$, $B$) are needed to generate radial profiles. The three constants defining an opacity approximation are a generalisation of the classical Shakura-Sunyaev models. 

\begin{figure}
    \resizebox{\hsize}{!}{\includegraphics[scale=0.8]{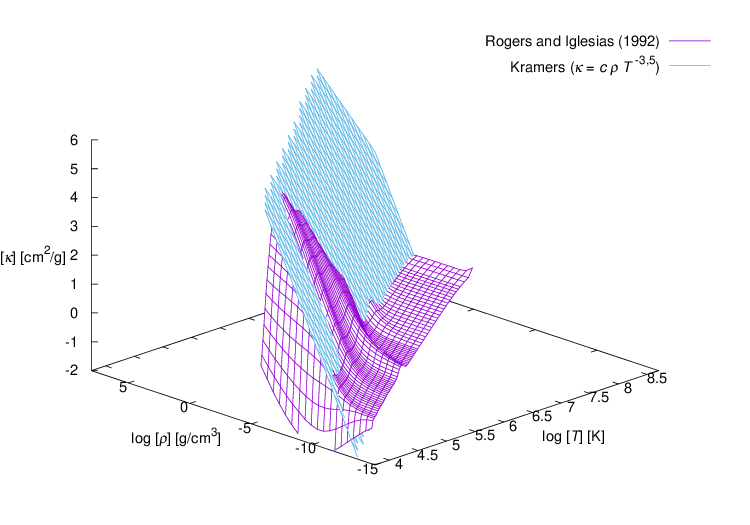}}
    \caption{\citet{1992ApJ...401..361R} 2D opacity function (magenta) approximated by a plane (in a logarithmic plot) in the interval corresponding to \kvthrd{Kramers} opacity (cyan).}
    \label{fig:rogandigl and krammer}
\end{figure}

We adopted the 2D opacity function $\kappa = \kappa(\rho,T)$ from \citet{1992ApJ...401..361R}. For defined regions, we found approximations in the form of Eq.~(\ref{eq:op_approx_general}). Initially, we focused on regions that were indicated as possibly corresponding to the disc discussed by \citet{2021A&A...645A..51B}, but as our investigation progressed we \kv{iterated our choices according to our intermediate results}. The approximations were optimised by converging parameters $\kappa_0$, $A$ and $B$ from Eq.~(\ref{eq:op_approx_general}) using the least-squares method
\mb{(as in Fig.~\ref{fig:rogandigl and krammer})}.

\subsection{\kvthrd{Scrutiny of the analytical models}} \label{sec:scrut}

\kvthrd{
We observe the following criteria, when judging the usefulness of the \mb{individual} analytical models.
\begin{itemize}
    \item We check the consistency of the calculated profiles with assumptions under which they were derived. Two types of possible inconsistencies are observed\mb{. First,} each class of models \mb{is} dependent on one of three assumptions about gas and radiation pressure. Second, the individual models are generated by a specific opacity approximation (through $\kappa_0$, $A$ and $B$), which are only valid in the \mb{specific part} of the $T-\rho$ space.
    \item In our models\mb{,} we treat \mb{$\alpha$} as a free parameter\mb{.} As we are aware of the uncertainty that goes along with \mb{our} assumptions, we prefer models that \mb{fulfil} the consistency discussed \mb{above} for a wide range of $\alpha$ values. We refer to the "stability" of the consistency.    
    \item We compare the values in the computed profiles with the observations and adjusted observations. This is again $\alpha$\mb{-}dependent. 
    \item The \citet{2021A&A...645A..51B} profiles are \mb{essentially} power laws. As noted in Sec.~\ref{sec:anal_models}, the modified Shakura-Sunyaev profiles are \mb{also approximated by} power laws \mb{for $r \gg 1$}. We \mb{thus} compare the exponents of the temperature profiles\mb{;} the observed temperature profile is proportional to $T \propto r^{-0.73}$ \citep{2021A&A...645A..51B}.
\end{itemize}
}

\subsection{\kvthrd{Time-dependent numerical simulations with general opacity tables}}

\kvthrd{\mb{A} final model of the \mb{$\beta$~Lyr~A} system is obtained by \mb{more} self-consistent numerical simulations, where the results from the analytical models are used as initial conditions. This \mb{choice} improves the chances of doing away with initial transients in the simulation as efficiently as possible. When we \mb{demonstrate} the internal consistency of the analytical models \mb{for specific parameters,} it \mb{also serves as} justification for the choice of included physics in the numerical model (e.g.\mb{,} the equation of state).}

\kvthrd{In these simulations, we do not assume a single \mb{power-law} opacity prescription in advance\mb{,} as is the case in the analytical models, instead we use \mb{more} general opacity tables (e.g., \citealt{2012ApJ...746..110Z,1994ApJ...427..987B}). The opacity table \mb{is} made up of \mb{several power-law} approximations\mb{,} with the advantage that an opacity prescription is chosen \mb{at each time step,} based on the \mb{local} conditions that arise \mb{during} the simulation (\mb{corresponding} $\rho$ and $T$).}

\kvfth{This step to  numerical simulations is necessary for a number of reasons}, \mbfth{namely, (i)}
\kvfth{the analytical models were derived assuming a steady state Keplerian disc, time-dependent simulations without an enforced velocity profile can test the validity of this assumption in the case of the studied system.
\mbfth{(ii)}~General opacity tables allow for the opacity regime to be \mbfth{coupled} to the physical state of the system.
\mbfth{(iii)}~The influence of phenomena like stellar irradiation that are not included in the analytical model \mbfth{should be also assessed.}
}

\kvthrd{The \mb{profiles} from this model are then used to interpret the observationally constrained profiles of \citet{2021A&A...645A..51B}.}

\section{Results} \label{sec: results}

\kvthrd{\mb{Hereinafter, we present} our main results from both the analytical and numerical models\mb{;} alternative models \kvfth{that we in the end rejected} are presented in \mb{Appendices~\ref{sec:other_anal} and \ref{sec:alter_sol}}. All presented analytical models are summarised in Tab.~\ref{tab:gasdommodel}, \mbfth{but} in the \mb{following} we will consider the models belonging to the \kvfth{class, which led to the best, self-consistent results} (i.e., gas pressure dominated, \kvfth{with Kramers opacity}).}

\subsection{Gas pressure dominated \kvthrd{analytical} models of the $\beta$~Lyr~A disc}

\begin{table*}
    \centering
    \caption{Overview of modified Shakura-Sunyaev models for $\beta$ Lyr A disc.}
    \begin{tabular}{c|c|c|c|c|c}
        Assumption & Designation & $\kappa(\rho, T)$ & Consistency & $T \propto r^{\gamma}$ & Figs. \\ \hline \hline
        $P_{\rm g} \gg P_{\rm r}$ & \kvthrd{Kramers} & $6 \cdot 10^{24} \rho T^{-\frac{7}{2}}$  & Yes & $-0.75$ & \ref{fig:rogandigl and krammer},\ref{fig:Pgas_Krammer_Sigma},\ref{fig:Pgas_Krammer_T},\ref{fig:Pgas_Krammer_P},\ref{fig:Pgas_Krammer_Aspect} \\
        $P_{\rm g} \gg P_{\rm r}$ & "Ridge" & $10^{7.67} \rho^{0.72} T^{-0.1}$ & Yes, except for $\alpha \leq 0.01$ & $-1.08$ & \ref{fig:Rho_opacity_approx_}, \ref{fig:Pgas_Rho_T}\\
        $P_{\rm g} \gg P_{\rm r}$ & High temperatures & $10^{18.6} \rho^{0.77} T^{-2.5}$ & Yes & $-0.79$ & \ref{fig:High temp_approx+opacity} \\
        $P_{\rm g} \approx P_{\rm r}$ & \kvthrd{Kramers} & $6 \cdot 10^{24} \rho T^{-\frac{7}{2}}$  & No & & \ref{fig:rogandigl and krammer} \\
        $P_{\rm g} \approx P_{\rm r}$ & "Ridge" & $10^{7.67} \rho^{0.72} T^{-0.1}$ & No & & \ref{fig:Rho_opacity_approx_},\ref{fig:Pap_Rho_P}, \ref{fig:PaP_Rho_T} \\
        $P_{\rm g} \approx P_{\rm r}$ & Inverse problem & $6 \cdot 10^{23} \rho^{2} T^{-1}$ & Only in the inner disc for $\alpha = 0.1$ & & \ref{fig:inverse_approx+opacity}, \ref{fig:PapP_inverse_Sigma}, \ref{fig:PapP_inverse_T}, \ref{fig:PapP_inverse_P}, \ref{fig:PapP_inverse_Aspect}  \\
        $P_{\rm g} \ll P_{\rm r}$ & \kvthrd{Kramers} & $6 \cdot 10^{24} \rho T^{-\frac{7}{2}}$  & No & & \ref{fig:rogandigl and krammer} \\
        $P_{\rm g} \ll P_{\rm r}$ & "Ridge" & $10^{7.67} \rho^{0.72} T^{-0.1}$ & No & & \ref{fig:Rho_opacity_approx_}  \\
        $P_{\rm g} \ll P_{\rm r}$ & Extreme temperatures & $6 \cdot 10^{23} \rho^{0.5} T^{-\frac{7}{2}}$ & Only close to $\alpha \approx 1.0$ & & \ref{fig:Extremetemp_approx+opacity}, \ref{fig:Prad_extrm_Sigma}, \ref{fig:Prad_extrm_T}, \ref{fig:Prad)_extrm_P}, \ref{fig:Prad_extrm_Aspect} \\
    \end{tabular}
    \tablefoot{
    The assumption refers to the choice of pressure approximation,
    the designation to the choice of opacity approximation (i.e., a region of the opacity function \citealt{1992ApJ...401..361R}) and $\kappa(\rho, T)$ is the specific form of the prescription.
    The consistency reports whether there are any contradictions
    between the computed profiles and assumptions used in the derivation.
    The power law is a qualitative description of the computed temperature profile 
    that dominates in the outer part of the disc,
    where the "bending" term (Eq.~(\ref{eq:Sigma and bending})) is negligible.}
    \label{tab:gasdommodel}
\end{table*}

Here, we apply gas pressure dominated profiles from Eqs.~(\ref{mod_shaksun_gas_start})-(\ref{mod_shaksun_gas_end}),
for \mbfth{a specific choice of the opacity prescription}.

\subsubsection{\kvthrd{Kramers opacity}}\label{sec:421}

\begin{figure*}
    \centering
    \begin{subfigure}{0.49\hsize}
    \resizebox{1.0\hsize}{!}{\includegraphics{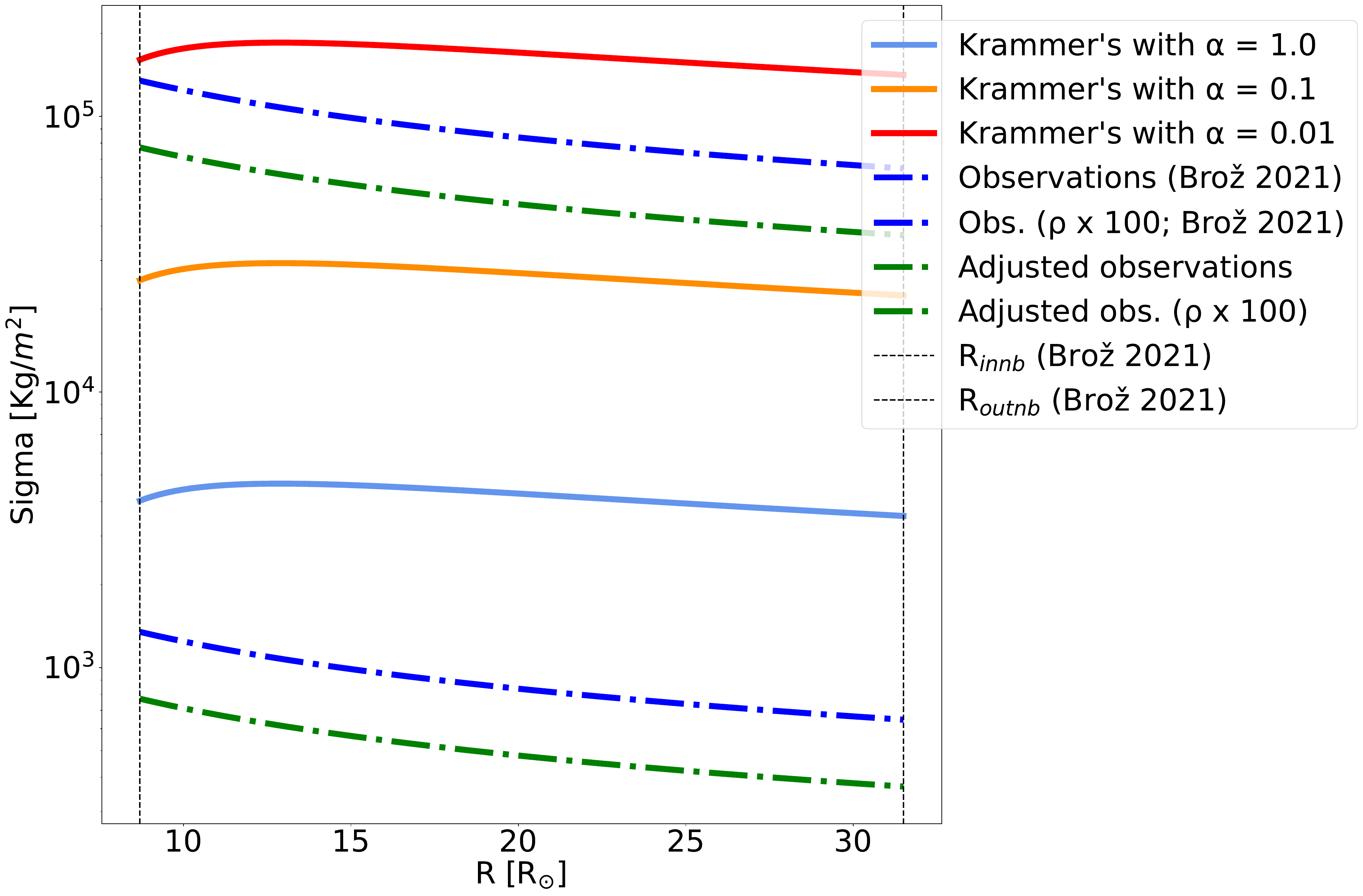}}
    \caption{Surface density $\Sigma$}
    \label{fig:Pgas_Krammer_Sigma}
\end{subfigure}
\hfill
\begin{subfigure}{0.49\hsize}
    \resizebox{1.0\hsize}{!}{\includegraphics{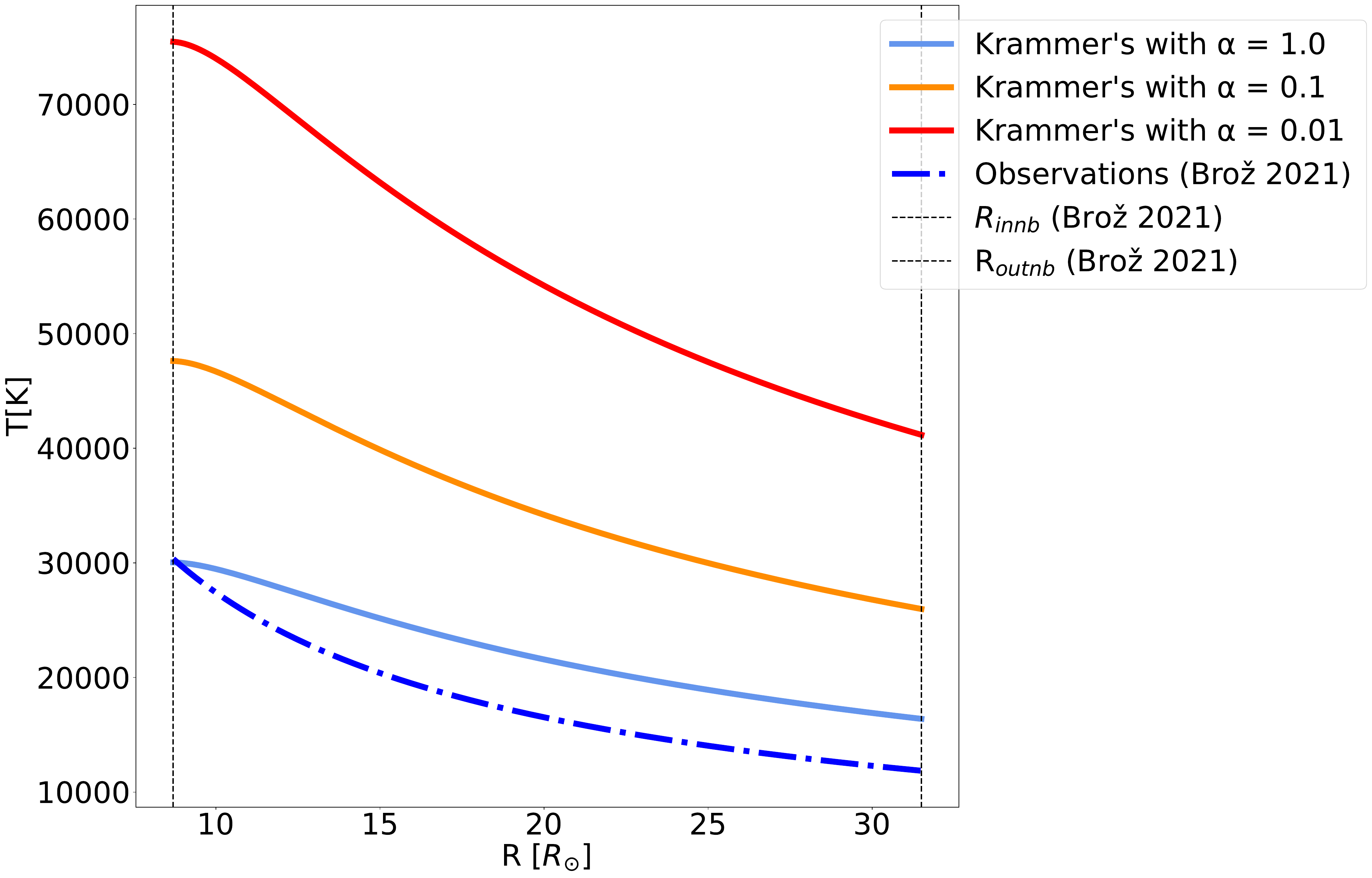}}
    \caption{Temperature $T$}
    \label{fig:Pgas_Krammer_T}
\end{subfigure}
\hfill
\begin{subfigure}{0.49\hsize}
    \resizebox{\hsize}{!}{\includegraphics{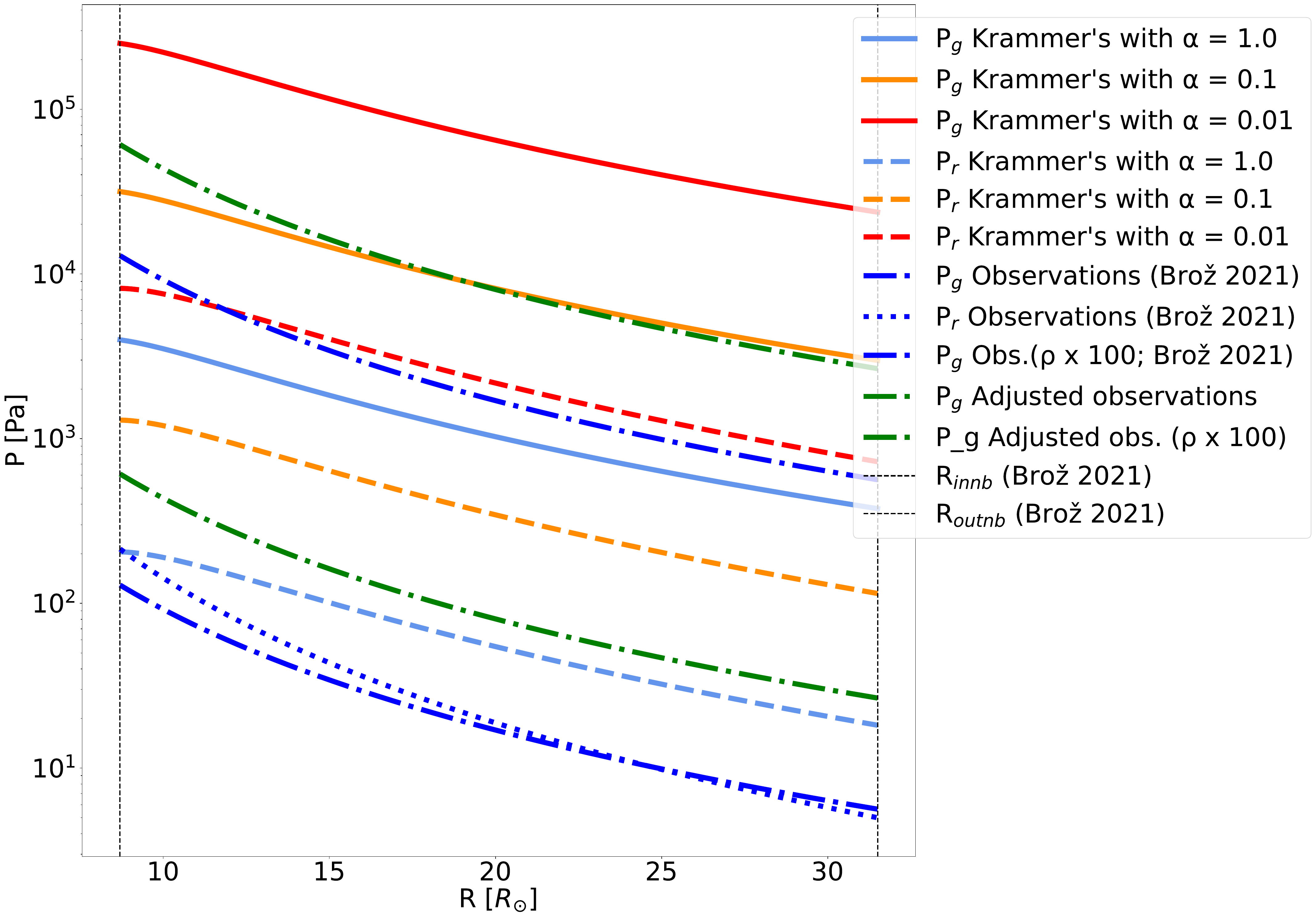}}
    \caption{Gas pressure $P_{g \rm}$ and radiation pressure $P_{r \rm}$}
    \label{fig:Pgas_Krammer_P}
\end{subfigure}
\hfill
\begin{subfigure}{0.49\hsize}
    \resizebox{\hsize}{!}{\includegraphics{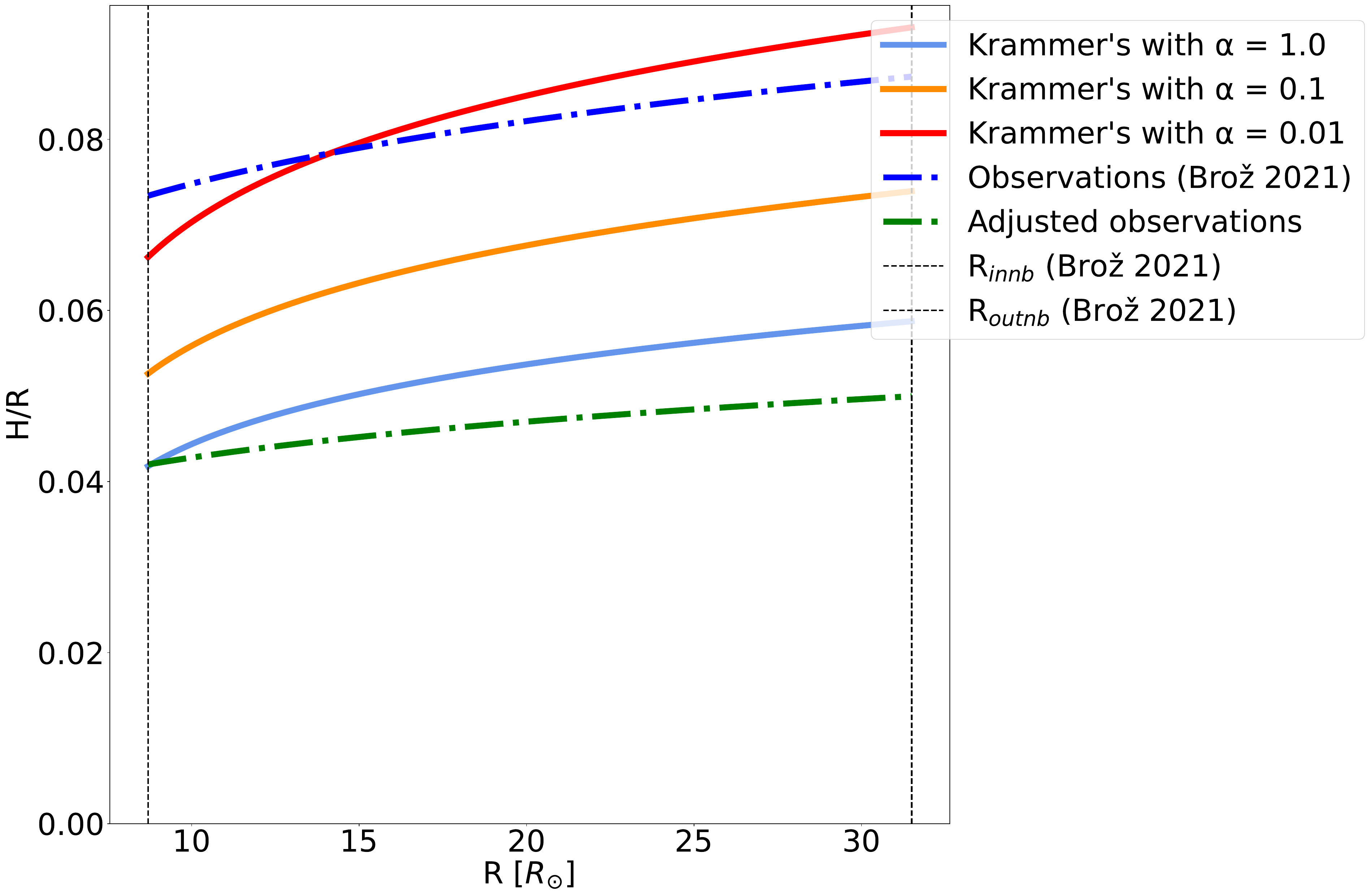}}
    \caption{Aspect ratio $H/r$}
    \label{fig:Pgas_Krammer_Aspect}
\end{subfigure}
\caption{Radial profiles of the surface density (a), the temperature (b), the gas and radiation pressures (c) and the aspect ratio of the $\beta$ Lyr\ae A disc for a modified Shakura-Sunyaev model, \kv{which is gas-pressure-dominated and with \kvthrd{Kramers opacity}}. The midplane temperatures and surface densities are significantly higher than the observed values, this generates also significantly higher pressure profiles. The observed aspect ratio is reached in hydrostatic equilibrium.}

\label{figs:Pgas_Krammer}
\end{figure*}

In Fig.~\ref{fig:rogandigl and krammer}, we plotted for illustration the 2D opacity function \citep{1992ApJ...401..361R} and the approximation of the region approximately corresponding to \kvthrd{Kramers} opacity (Eq.~(\ref{eq:krammer})), which is valid for a wide range of temperatures.
\begin{equation} \label{eq:krammer}
     \kappa = 6 \cdot 10^{24} \rho T^{-\frac{7}{2}}\,.
\end{equation}
The resulting hydrodynamical profiles are plotted in Fig.~\ref{figs:Pgas_Krammer}.

The profiles generated by the gas pressure dominated model with \kvthrd{Kramers} opacity are consistent with the assumptions under witch the model was derived for all tested values of the $\alpha$ parameter. 

The temperature profiles generated by the \kvthrd{Kramers} model (for all $\alpha$ parameters) are significantly higher than in the observed profiles. The analytic model generates a hotter disc with only viscous dissipation, while in reality, stellar irradiation will serve as an additional energy source. The case of the theoretical limit $\alpha = 1.0$ \citep{1973A&A....24..337S} approaches the observed profile, but the agreement is still not satisfactory.

\kvthrd{An inspection of Tab.~\ref{tab:gasdommodel} shows \mb{that} this model produces qualitatively most compatible temperature profile, $T \propto r^{-0.75}$}. (This similarity was also hinted at by \citet{2021A&A...645A..51B}.)

The model shows significantly higher surface density profiles $\Sigma$, for $\alpha = 0.1$ and $\alpha = 0.01$ the values of the $\Sigma$ profiles are over 10 and 100 times larger than the observations. (For reference, in Fig.~\ref{fig:Pgas_Krammer_Sigma} we plotted $\Sigma$ profiles for the observations and adjusted observations, multiplied by a factor of 100.) Even for the theoretical limit of $\alpha = 1.0$, it is still about 4 times larger. 

\kvthrd{We} show the pressure profiles in Fig.~\ref{fig:Pgas_Krammer_P}. Due to the high temperatures and densities in the disc, both gas pressure and radiation pressure values are significantly higher than can be calculated from the observed profiles (For reference, in Fig.~\ref{fig:Pgas_Krammer_P} we plotted gas pressure profiles calculated from the observations and adjusted observations assuming a 100 times larger density.)

\kvthrd{Kramers} model generates aspect ratios (for $\alpha = 1.0,0.1,0.01$) between the observations and adjusted observations, as demonstrated in Fig.~\ref{fig:Pgas_Krammer_Aspect}. This is due to the high midplane temperatures, allowing the $H$ profile given in \citet{2021A&A...645A..51B} to be reached in a hydrostatic equilibrium.

\subsubsection{Implications from \mbfth{all} analytical models} \label{sec:ancol}

\kvthrd{From the application of the modified Shakura-Sunyaev models \mbfth{(Tab.~\ref{tab:gasdommodel})} we can already draw several conclusions.}

\kvthrd{The models based on assuming gas pressure dominance are robustly consistent, whilst other classes of models are either contradictory to their own assumptions or their consistency is unstable. We take this to be an argument that gas pressure plays a dominant role in the disc.}

\kvthrd{None of the models match the densities or temperatures of the observationally constrained profiles well\mb{,} without choosing \mb{the} extreme case of $\alpha = 1.0$. The generated densities and temperatures tend to be significantly higher.}

\kvthrd{Considering the criteria \mb{from} Sec.~\ref{sec:scrut}\mb{,} we see \mb{the} gas pressure dominated disc with \mb{$\alpha = 0.1$ and} \kvthrd{Kramers} opacity as the preferred model.}

\subsection{\kvthrd{Time-dependent model} with the \cite{2012ApJ...746..110Z} opacity table} \label{sec:52}

\begin{figure}[htbp!]
    \centering
    \includegraphics[width=9cm]{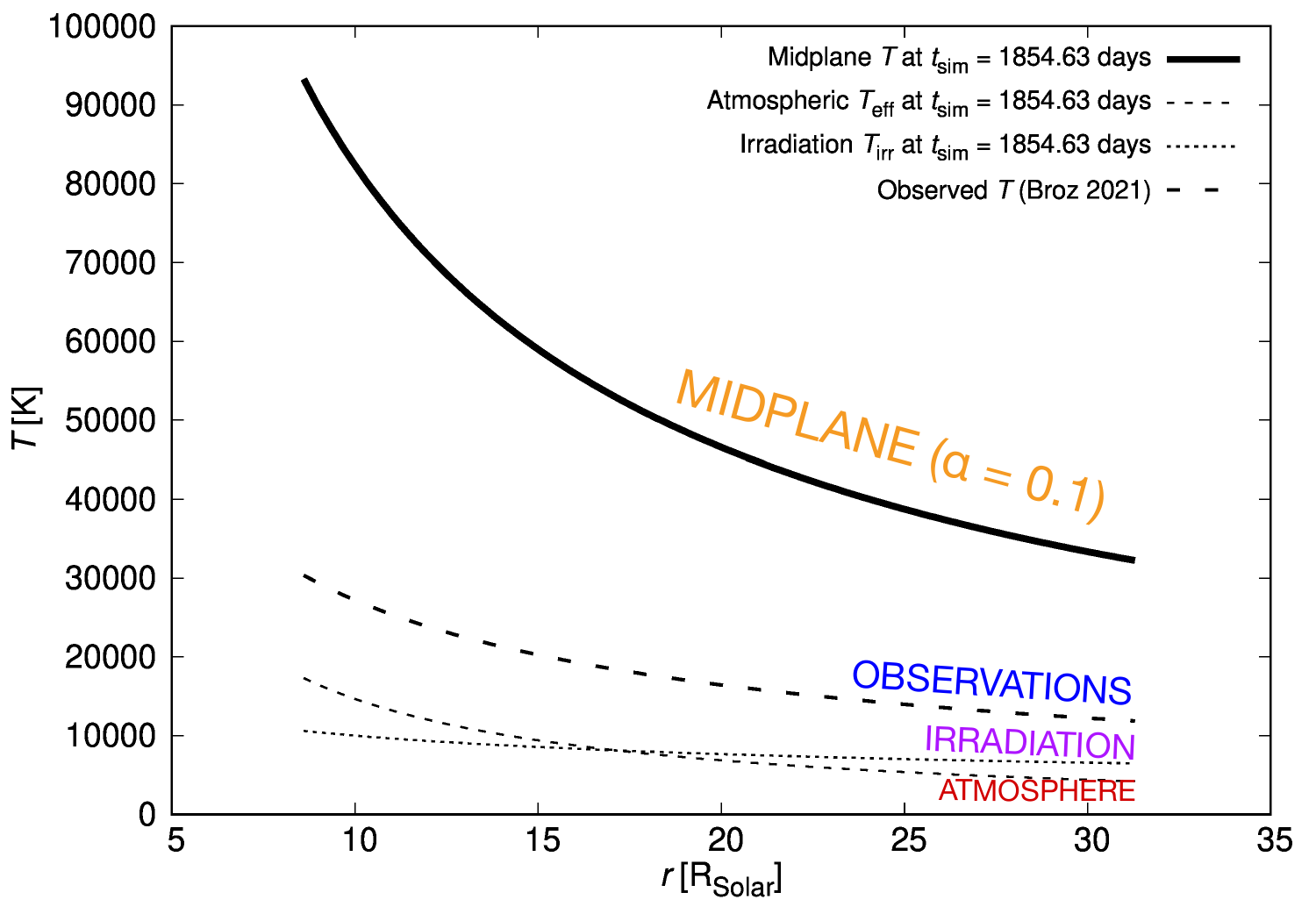}
    \includegraphics[width=9cm]{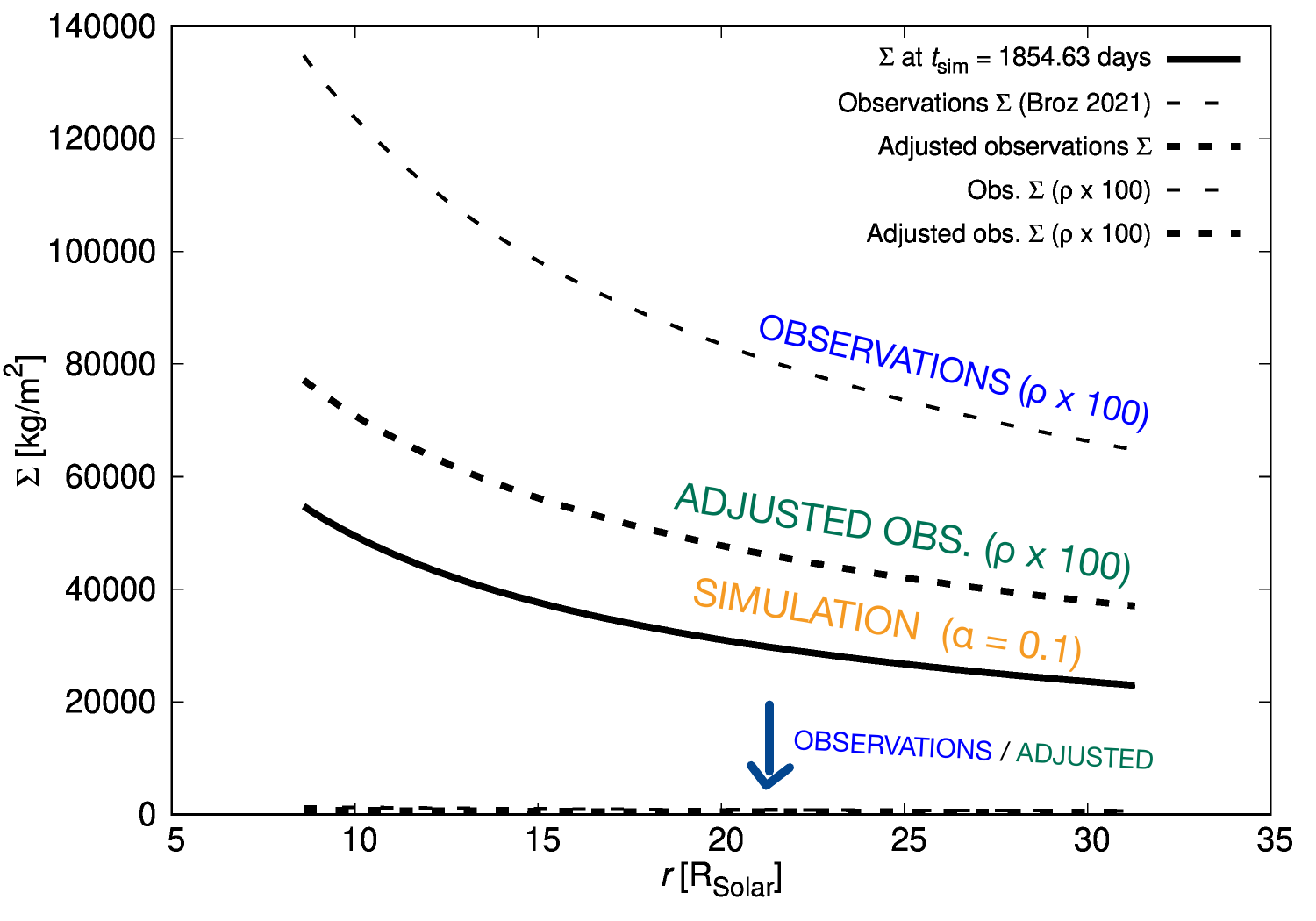}
    \includegraphics[width=9cm]{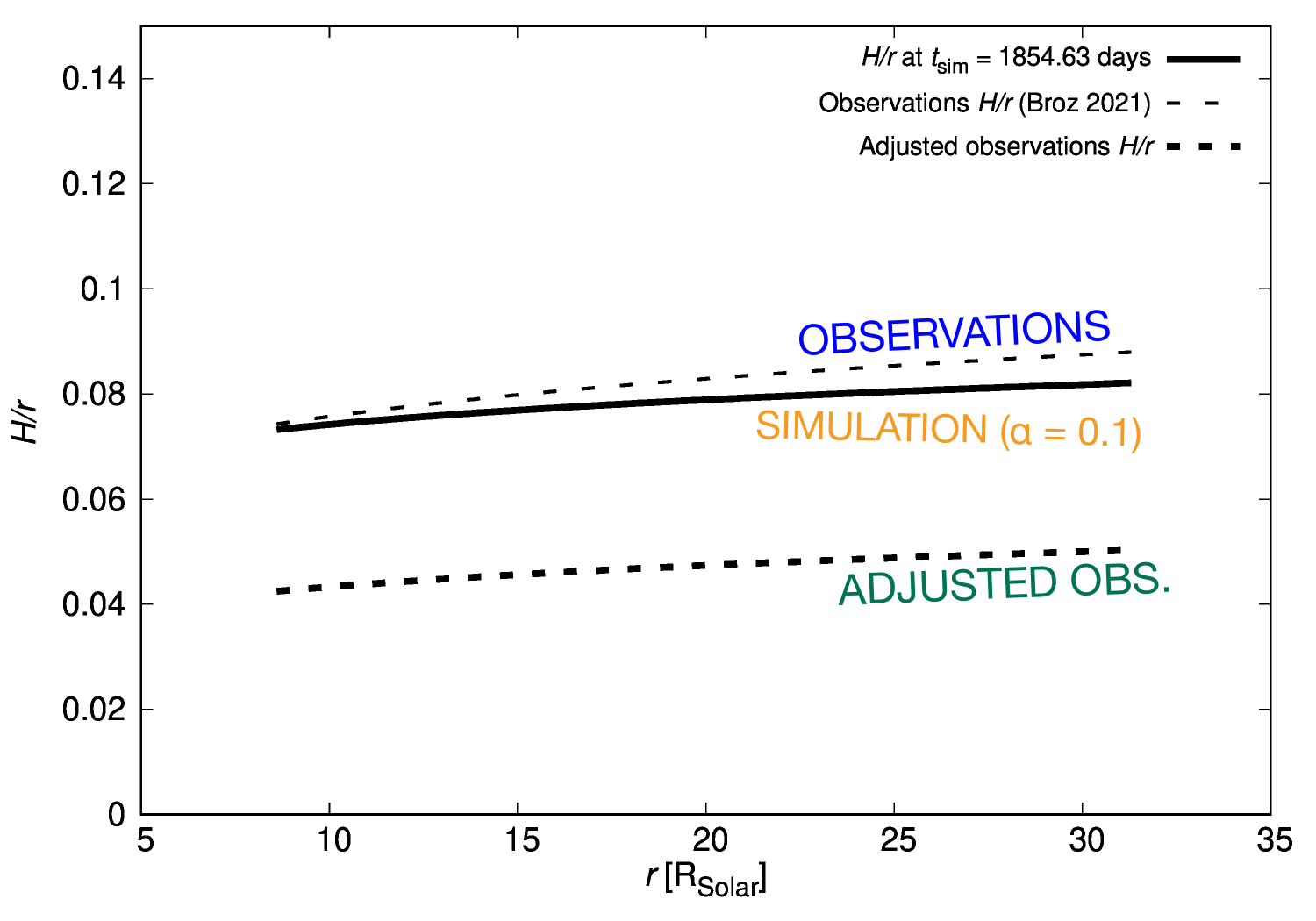}
    \caption{
    Temperature $T(r)$ (top),
    surface density $\Sigma(r)$ (middle),
    and aspect ratio~$H(r)$ (bottom) profiles
    of the time-dependent numerical model (from Tab. \ref{tab:fargo_input_zhu_gen} with the \citet{2012ApJ...746..110Z} general opacity table).
    The midplane temperatures are significantly higher than the observations. 
    The calculated atmospheric temperatures $T_{\rm eff}$ and irradiation temperatures $T_{\rm irr}$ are of comparable magnitudes to the observations, but the correspondence between the temperature profiles isn't satisfactory.
    Surface densities from the model are almost 100 times the observed value.
    The observed aspect ratio is reached in hydrostatic equilibrium.
    }
    \label{fig:zhu_gen}
\end{figure}

\mb{For the preferred analytical model
(gas-pressure-dominated, \kvthrd{Kramers} opacity, $\alpha = 0.1$),
we constructed a corresponding numerical model.}
\mb{However}, we used \mb{the} general opacity table by \cite{2012ApJ...746..110Z}
and input parameters given in Tab.~\ref{tab:fargo_input_zhu_gen}.
\mb{Our} results are presented in Fig.~\ref{fig:zhu_gen}.

The disc reached a steady state within one year of simulation time
and remained without temporal evolution for the rest of the simulation time
(the run was terminated after $2\cdot 10^3$ days).
We calculated the viscous timescale profile
and found a range from about $t_{\nu \rm in} = 0.5$ years in the inner part of the disc
to about $t_{\nu \rm out} = 3$ years in the outer part.

According to the opacity table,
the values of $\kappa_{\rm in} \approx 10^0 \ \rm cm^{-2}\,g^{-1}$ in the inner part of the disc and $\kappa_{\rm out} \approx 10^1 \ \rm cm^{-2}g^{-1}$ in the outer part were reached, without any opacity transitions in the radial profile.
This closely resembles a profile generated by a simulation with the prescription $\kappa = 10^{18.6} \rho^{0.77}T^{-2.5}$,
which was previously used for one of the gas-pressure-dominated analytical models ("High temperatures" model)\kvsec{, we noted that it should be considered a slight variation of the \kvthrd{Kramers} opacity regime.}

The disc rotates with a slightly sub-Keplerian velocity due to a radial pressure gradient in the disc,
however, the deviations are minor.
The radial velocities are dependent on the efficiency of the angular momentum redistribution
and are coupled to the $\Sigma$ radial profile via the fixed mass transfer.
Radial velocities in the disc range from about $600\,{\rm m}\,{\rm s}^{-1}$ on the inner boundary
to 400 m$\cdot$s$^{-1}$ on the outer boundary.

Due to the high temperatures and densities, the model can reach a good agreement with the observed aspect ratio,
even under the assumption of a vertical hydrostatic equilibrium.

\paragraph{The $\Sigma$ problem.}
Comparing the surface density profile and the \citet{2021A&A...645A..51B} \kvthrd{observationally constrained profile},
the densities from our numerical simulation reach values almost 100 times larger. 
This discrepancy in the profiles of surface density is still acceptable.
As discussed in Sect.~\ref{sec:obs_param},
the observed $\Sigma$ profile is a lower limit to densities that could explain the observations.
Taking dynamics into consideration, we require a fixed amount of matter
that must be transferred through the disc (determined by $\dot{M}$, $v_r$, and $\alpha$).

\paragraph{The temperature problem.}
The midplane temperature of the disc from our numerical simulation
is around three times higher than the observed profile of \citet{2021A&A...645A..51B},
\mb{which} is the most dramatic discrepancy.
An important aspect to consider is that each approach has completely different assumptions
about the vertical temperature profile of the disc. Let us discuss each in turn:
\begin{itemize}
    \item \citet{2021A&A...645A..51B} assumed a vertically isothermal atmosphere. The disc is optically thick, hence the observations by which they constrain their model are observations of the atmosphere. Inevitably, they infer the midplane temperature as being equal to the atmospheric temperature. They also allow for a temperature inversion in their model, so that the outer atmosphere can even be hotter than the midplane temperature.
    
    \item \citet{2017A&A...606A.114C} assumed the atmosphere model derived by \citet{1990ApJ...351..632H}. The atmosphere dims the midplane and the face of the disc radiates with an atmospheric temperature (Eq.~(\ref{eq:q_{vert}})), which is lower.
\end{itemize}
If we reject the assumption of an isothermal atmosphere
and re-interpret the \kvthrd{temperature p}rofile of \citet{2021A&A...645A..51B}
as the measured temperature of the atmosphere,
then it should be compared to the atmospheric temperature $T_{\rm eff}$
inferred from the midplane temperature according to \citet{1990ApJ...351..632H},
\begin{equation}
        T_{\rm eff} = \frac{T}{(\tau_{\rm eff})^{\frac{1}{4}}} \, .
\end{equation}
Further, a theoretical upper thermodynamic limit $T_{\rm eff}$ could be the irradiation temperature $T_{\rm irr}$.
Even though $T_{\rm eff}$ and $T_{\rm irr}$ are of comparable magnitudes,
the agreement is still poor.
The atmospheric temperature profile $T_{\rm eff}$ is now lower than the observed profile
and surpasses the limiting $T_{\rm irr}$ profile in the inner part of the disc.

\paragraph{Solution 1: Temperature inversion in the vertical stratification}

\begin{figure}[htbp!]
    \centering
    \includegraphics[scale=0.15]{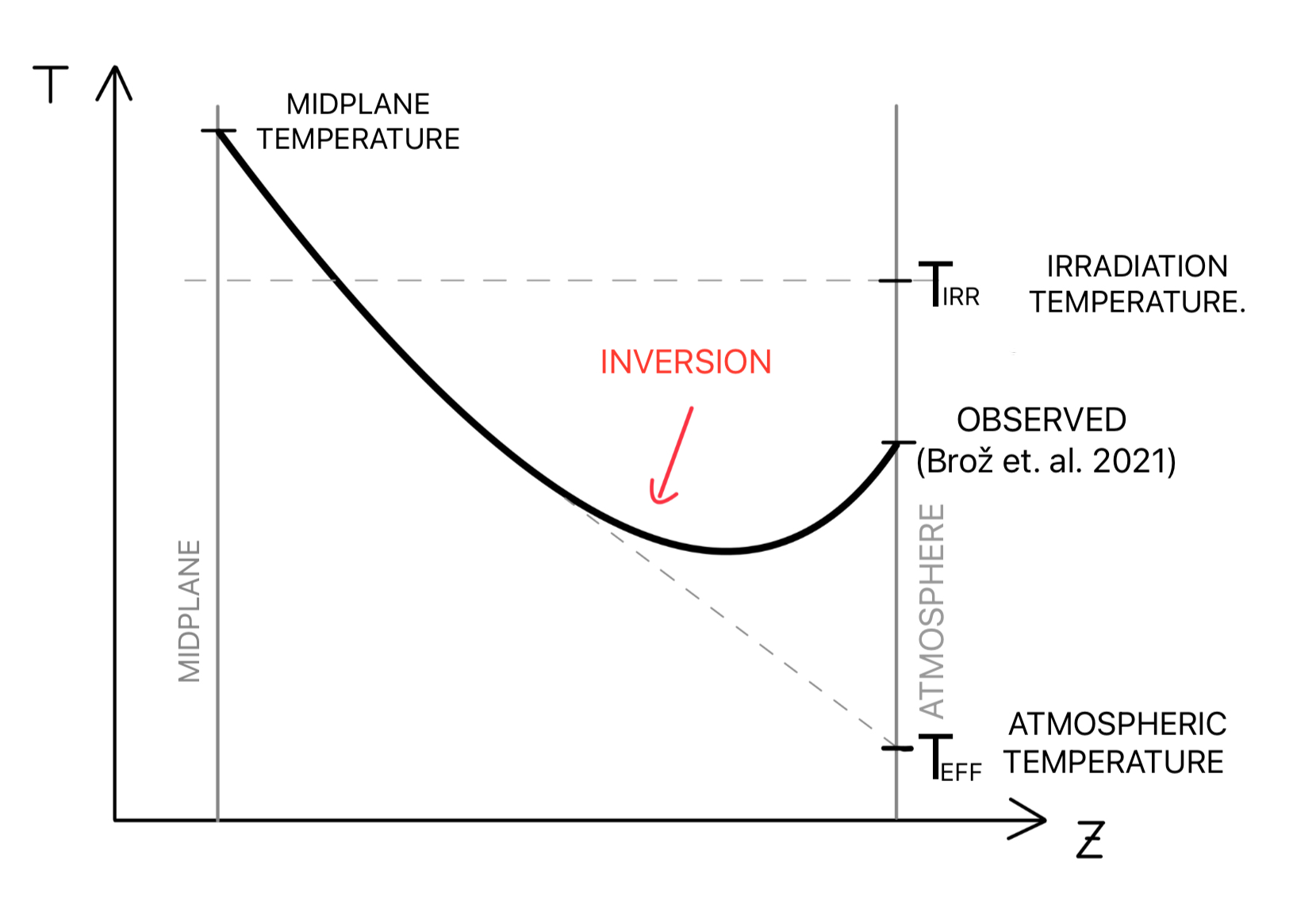}
    \caption{
    A sketch of how the presence of a temperature inversion in the vertical profile of the disc could reconcile the computed midplane temperature $T_{\rm mid}$ profile, the observed temperature profile and the calculated atmospheric $T_{\rm eff}$ and irradiation $T_{\rm irr}$ profiles. 
    }
    \label{fig:diagram}
\end{figure}

\kvthrd{These further discrepancies \mb{point} to a more complex vertical temperature profile. In \citet{2021A&A...645A..51B}, an inversion in the vertical temperature profile was considered.
In Fig.~\ref{fig:diagram}, we demonstrate how an inversion, e.g., due to non-thermal irradiation,
could explain the discrepancies between temperature profiles from the observations and from our numerical models.
With increasing distance from the midplane, the temperature decreases approximately according to the \citet{1990ApJ...351..632H} model.
At the level of the inversion, the temperature starts growing again,
up to where the matter becomes optically thin,
where it reaches the observed value. Similar temperature inversions can be seen in models of other Be stars (e.g., in decretion discs; \citealt{2021ApJ...922..148G}).}

\kvthrd{Alternative solutions are discussed in Appendix~\ref{sec:alter_sol}.}

\begin{table}
\caption{Time-dependent numerical model input parameters.}
\centering
\begin{tabular}{lcc}
\hline \hline Parameter & Unit & Value \\
\hline
$\Sigma_{0}$                 & $\rm kg \cdot m^{-2}$ & 26332.9 \\
$\sigma_{0}$                 & 1 & 0.42 \\
$\left(H/r\right)_0$         & 1 & 0.05 \\
$\kappa$                     & $\rm cm^{2} \cdot g^{-1} $ & \citet{2012ApJ...746..110Z} \\
$C_{\rm k}$                  & 1 & 0.6 \\
$\alpha$                     & 1 & 0.1 \\
$\gamma$                     & 1 & 1.4 \\
$\mu$                        & $m_{\rm u}$ & 0.5 \\
$A$                          & 1 & 0.5 \\
$r_{\rm sm}$                 & $R_{\rm HILL}$ & 0.6 \\
$t_{\rm total}$              & $\rm yr$ & 5 \\
$N_{\text{r}}$               & 1 & 200 \\
$N_{\phi}$                   & 1 & 2 \\
\hline
\end{tabular}
\label{tab:fargo_input_zhu_gen}
\tablefoot{For full definitions of individual parameters, see \citet{2017A&A...606A.114C}.}
\end{table}

\section{Discussion} \label{discussion}

Finally, we discuss more general implications of our modelling of \mb{the} $\beta$~Lyr~A \mb{disc}.

\paragraph{Viscosity:} Overall \kvthrd{our preferred (time-dependent) model was} obtained for $\alpha \approx 0.1$; where we tested different orders of magnitude of $\alpha$ and did not perform any "fine-tuning". This value is consistent with evidence for fully ionised thin accretion discs (\kvthrd{values in literature given at 0.1-0.3\mb{;} }\citet{alphabig,2018MNRAS.479.2214G,2021ApJ...922..148G}).
The \kvthrd{model of the} limit case of $\alpha = 1.0$ was also considered as a possible way to reconcile models and observations (Appendix~\ref{high_alpha}), but it is not preferred, because without a clear idea of how this value could be obtained from the microphysics of viscosity or higher-dimensional macroscopic phenomena in the disc \citep{SpiralShocksTheo}, it is somewhat of an arbitrary choice without objective justification. \kv{Possible dynamical phenomena contributing negative angular momentum such as spiral waves must be explored with 2D and 3D simulations.}
In all our simulations, we always used $\alpha$ which is constant with the radial distance. In principle, $\alpha$ may vary across the disc, e.g., \citet{2016ApJ...827..144F} propose a formula that increases $\alpha$ in zones where the gas is hot enough to be sufficiently ionised for the magneto-rotational instability to occur. However, $\beta$~Lyr~A is hot enough to be ionised everywhere (we get a constant $\alpha$ again). No quantities show any 'jumps' or regime transitions, hence only some weak power laws may be considered. The effect on the result likely won't be dramatic.
\kvthrd{The influence of $\alpha$ on the equilibrium state can be estimated from \mb{our modified} Shakura-Sunyaev models, e.g.\mb{,} for the temperature profile in the case of a gas pressure dominated disc with \kvthrd{Kramers} opacity\mb{,} $T \propto \alpha^{-\frac{1}{5}}$}.

\paragraph{Conservation of mass:} One of the key parameters of our model is the fixed mass transfer rate $\dot M$, the value of which was inferred from the period change of the binary, assuming mass conservation. It is to be noted that in reality, the conservation of mass is imperfect. Accretion of angular momentum from the disc can be at the origin of the rotational velocities of Be stars. The spin-up of the gainer can lead to super-synchronous rotation which modifies the Roche equipotentials, where the lobe moves closer to the star \citep{1958LIACo...8..411P}. This implies more likely overflows on the gainer's side and hence non-conservative mass transfer \citep{2024ApJ...966L...7L}. Documented jets and the optically thin material surrounding the system are further indications of some mass loss \citep{1996A&A...312..879H,2021A&A...645A..51B,2023MNRAS.525.5121V}. \kv{The value of $\dot{M}$ has a significant influence on the models (e.g., for the \kvthrd{gas} pressure dominated analytical model assuming \kvthrd{Kramers} opacity, $\Sigma \propto \dot{M}^{\frac{7}{10}}$), hence the amount of optically thin matter around the system needs to be constrained.} 

\paragraph{Timescales:} In our time-dependent models, the disc relaxed quite quickly and remained so for the rest of the simulation time (a multiple of viscous timescales \citep{Pringle_1981ARA&A..19..137P}; $t_{\nu} \approx 0.5 \, \text{yrs}$ at the inner boundary to $t_{\nu} \approx 3\, \text{yrs}$ at the outer boundary). A steady-state configuration therefore seems a probable configuration for the disc. \kvthrd{Let us point out, that also in the presented alternative models a steady state was reached.}

\paragraph{Orbital velocities:} The radial pressure gradient in our numerical simulations resulted in only a slightly sub-Keplerian rotation. For many practical purposes, it is still valid to approximate the disc as Keplerian. Considering also the reached steady state, the results of the time-dependent simulations fulfil the assumptions of the applied analytical models \citep{1973A&A....24..337S}. This is in agreement with the broad consensus that Be stars are surrounded Keplerian discs (see the review by \citet{2013A&ARv..21...69R}). \kvthrd{Further, there is a general agreement between our preferred analytical and numerical models, which is encouraging for the application of equilibrium models.}

\paragraph{Opacity:} The \kvthrd{preferred} analytical models were obtained for \kvthrd{Kramers} opacity;
a more general opacity table by \citet{2012ApJ...746..110Z} also produced an opacity profile \kvsec{with values} close to the \kvthrd{Kramers} opacity.
The disc exhibits no opacity transitions. The viscosity parameter $\alpha \approx 0.1$ preferred in this study is common in ionised accretion discs, hence a combination of free-free (bremsstrahlung) and bound-free (photoionisation) absorption 
--both approximated by \kvthrd{Kramers} opacity law--
is a corresponding candidate for the main opacity source \citep{zel2012physics}. This is consistent with the density ($\rho \approx 10^{-5}$\,kg$\cdot$m$^{-3}$) and temperature $T \approx 10^{4} - 10^{5}$\,K) profiles in our model of the disc interior (Fig.~\ref{fig:zhu_gen}), for which the Saha equation predicts the ionisation fraction safely above 0.5.

Moreover, the vertical opacity profile has a substantial impact on the disc \kvthrd{(Appendix~\ref{vert_op_prof})} and should be considered with care.
Here, we assumed the same opacity regime within the whole vertical extent,
but we introduced the integrated effect of the $\kappa(z)$ dependency through the $C_{\rm k}$ factor in calculating optical depth (Eq.~(\ref{eq:tau_opt})).
\kv{Yet, for a more precise result, the vertical opacity profile needs to be modelled self-consistently, hand in hand with the vertical temperature profile.}

\paragraph{Pressure:} The analytical models are consistent for a wide range of opacities and values of $\alpha$ with their own assumptions only when $P_{\rm g} \gg P_{\rm r}$ is assumed. The (unmodified) numerical model \citep{2017A&A...606A.114C} contains an ideal-gas equation of state (Eq.~(\ref{eq:fargoEqofSt})), so it is based on the same assumption. We also considered an additional increase due to radiation pressure, but its influence is indeed minor.

\paragraph{Surface densities and radial velocities:} Due to the fixed mass transfer rate, the surface density profile and radial velocity profile are strongly coupled in our models. Radial velocities in the disc are of the order of $10^{2}\,{\rm m}\,{\rm s}^{-1}$. In both the analytical and numerical models (computed for $\alpha = 0.1$), we get surface densities of the order of $10^4\,{\rm kg}\,{\rm m}^{-2}$ throughout the disc. This is close to 100 times the observed value (which is only a lower limit to the real densities of the optically thick matter; \citealt{2021A&A...645A..51B})\mb{.}

\kvthrd{An improved model that would consider non-conser\-vative mass transfer (hence lower $\dot{M}$, \mb{while keeping~$\dot P$}) would allow for both lower surface densities and lower radial velocities without a significant influence on the observables of the disc.}

\paragraph{Temperature profile:} Both our \kvthrd{preferred} models (for $\alpha = 0.1$) indicate high midplane temperatures, close to $T \approx 10^5\,\text{K}$ at the inner boundary to $T \approx 3 \cdot 10^4\,\text{K}$ at the outer boundary. This is significantly higher than the observed temperatures of \cite{2021A&A...645A..51B}. In fact, the latter values were derived \kvthrd{from fitting observations} of the radiative atmosphere of the disc (at the boundary of the optically thick matter). When we computed the atmospheric temperatures from our \kvthrd{numerical} model (under the assumption of the atmosphere by \citet{1990ApJ...351..632H}),
\kvthrd{\mb{considered the possibility of}} temperature inversion in the upper atmosphere,
we obtained the magnitudes comparable to the observations (i.e., $3 \cdot 10^4$ to $1 \cdot 10^4 \, \text{K}$).
The inversion is consistent with the strong $H_{\alpha}$ emission.
Temperature inversions are features seen in other Be star models, e.g., in \citet{2021ApJ...922..148G}. 
\kv{We thus emphasise the need for higher-dimensional models also in the case of $\beta$~Lyr~A.}

Studies of similar systems report lower temperatures; \citet{Mennickent_2021A&A...653A..89M} observations of a $\beta$-Lyr\ae-type binary shows disc temperatures of the order of $10^3\,\text{K}$.
However, they also demonstrate that the temperature has a positive correlation with the mass transfer rate, in agreement with our analytical model.
The lower temperatures are thus arising due to the lower mass of the binary
(4.83 + 1.06$\,M_\odot$).
Similarly, the \citet{2021ApJ...922..148G} model of the decretion disc of the Be shell star 1~Del shows temperatures of the order of $10^3\,\text{K}$, even in the hottest parts of the interior. This is because the disc has densities 1000 times lower and radial outflow velocities 10-100 slower than in our model of $\beta$~Lyr~A.  

\kvthrd{As an alternative to the presence of an inversion, the possibility of vertical convective energy transport should be explored. This has been shown as a possibility in circumstellar discs \citep{2012A&A...539A..18M}.
\mb{If} present \mb{in the $\beta$~Lyr~A disc}, it would constitute a more effective cooling mechanism.
The more gradual temperature gradient could allow for lower midplane temperatures\mb{,}
while keeping similar atmospheric temperatures.}

\paragraph{Aspect ratio:} Due to the high midplane temperatures and densities, the observed aspect ratio of $H/r \approx 0.08$ can now be reached in a hydrostatic equilibrium. 
\kvthrd{The sensitivity of \mb{$H/r$} on the \mb{individual} input parameters can be \mb{also inferred} from \mb{our} analytical models.}
Moreover, if the contribution of radiation pressure is considered, the vertical extent of the disc further increases \citep{2021ApJ...910...31M}\mb{.}

\section{Conclusions}\label{sec:conslusions}

We presented \kv{analytical and numerical} hydrodynamical models of the circumstellar disc around the gainer of the $\beta$~Lyr~A system at the phase of rapid mass transfer.
The consistency check of the analytical models points to gas pressure being dominant in the disc; the contribution of the radiation pressure could be at most ${\sim}\,10\%$.
We estimated the viscosity parameter governing the angular momentum redistribution to be close to $\alpha \approx 0.1$. This is within the range of typical values of highly ionised accretion discs.
\kv{Both the analytical models and the general opacity tables indicate that the} dominant source of opacity in the optically thick disc is free-free and bound-free absorption as modelled by \kvthrd{Kramers} opacity
(see Sec.~\ref{sec:421} for details).

All our time-dependent models show the disc is in a steady state with only slightly sub-Keplerian rotational velocities.
\kv{In our model, the} mass transfer rate of $\dot M = 2\times 10^{-5}\,M_\odot\,{\rm yr}^{-1}$ is achieved with the surface densities ranging from 20000 to 60000$\,{\rm kg}\,{\rm m}^{-2}$ and the radial velocities of the order of 100$\,{\rm m}\,{\rm s}^{-1}$.
Temperatures in the midplane reach $30000$ to over $90000\,\text{K}$; the respective Shakura-Sunyaev models predict a power law exponent of $-0.75$.
\kv{Given the remaining `tension' between the density and temperature profiles
(discussed in Sec.~\ref{sec:52}),
we predict that} the vertical temperature profile is likely complex with a temperature inversion in the top layers.
\kv{A comparison to \kvthrd{fitted} observations shows that vertical} hydrostatic equilibrium can be preserved with an aspect ratio $H/r$ of 0.08.
On top of the steady state, \kv{timescales comparison shows} there is a possibility of spiral waves, which would transport a negative angular momentum to the inner parts of the disc.

\paragraph{Future work:}
As a continuation of this work, we propose to study the $\beta$~Lyr\ae~A system
assuming a non-conservative mass transfer
and to discuss a possibility of temporal evolution of the system's intrinsic parameters
(similarly as in \citet{Mennickent_2021A&A...653A..89M}).
Several variations of the numerical model were suggested to address the discrepancy in the temperature profile:
a convective instability,
a temperature inversion in the vertical direction,
as well as edge-on irradiation from the central star.
Other phenomena such as spiral waves should be studied with 2D and 3D simulations.

\begin{acknowledgements}
M.B. was supported by the Czech Science Foundation grant GA21-11058S.
K.V. is a fellow of the International Max Planck Research School for Astronomy and Cosmic Physics at the University of Heidelberg (IMPRS-HD) and acknowledges financial support from IMPRS-HD.
During revisions K.V. was supported by the Klaus Tschira Foundation.
\end{acknowledgements}

\bibliographystyle{aa}
\bibliography{biblio}

\begin{appendix}
\section{\mbfth{Fluid equations in numerical model}}
\label{sec:hydroeq}

\mbfth{The equations in our time-dependent, 2D, numerical model are as summarised follows \citep{2000A&AS..141..165M,2017A&A...606A.114C}:}
\begin{equation}
\frac{\partial \Sigma}{\partial t}+\nabla \cdot(\Sigma v)=0 \, ,
\end{equation}
\begin{equation} \label{eq:eqofmotionFARGO}
\frac{\partial \boldsymbol{v}}{\partial t}+\boldsymbol{v} \cdot \nabla \boldsymbol{v}=-\frac{1}{\Sigma} \nabla P+\frac{1}{\Sigma} \nabla \cdot \boldsymbol{\pi}-\frac{\int\rho \nabla \Phi \mathrm{d} z}{\Sigma} \, ,
\end{equation}
\begin{equation} \label{eq:energyFARGO}
\frac{\partial \epsilon}{\partial t}+\nabla \cdot(\epsilon \boldsymbol{v})=-P \nabla \cdot \boldsymbol{v}+Q_{\mathrm{visc}}+Q_{\mathrm{irr}}-Q_{\mathrm{rad}} \, ,
\end{equation}
\begin{equation} \label{eq:fargoEqofSt}
P = \Sigma \frac{\mathcal{R}T}{\mu} = (\gamma - 1) \epsilon  \, ,
\end{equation}
where
$t$ is the time,
$\Sigma$ the surface density $\rho$,
$\boldsymbol{v}$ the vertically integrated velocity (a 2D vector: $\boldsymbol{v}=(v_{r},v_{\phi})$),
$P$ the vertically integrated pressure,
$\boldsymbol{\pi}$ the viscous stress tensor,
$\Phi$ the gravitational potential,
$\epsilon$ the vertically integrated internal energy,
$T$ the midplane temperature of gas,
$Q_{\textrm{visc}}$ the viscous heating term,
$Q_{\textrm{irr}}$ the stellar irradiation term, and
$Q_{\textrm{rad}}$ the radiative diffusion term.

\section{Other explored analytical models} \label{sec:other_anal}

\subsection{\kvfth{Gas pressure dominated models of the $\beta$~Lyr~A disc}}

\subsubsection{"Ridge" opacity}
\begin{figure*}[h]
    \centering
    \begin{subfigure}{0.46\hsize}
    \resizebox{1.0\hsize}{!}{\includegraphics{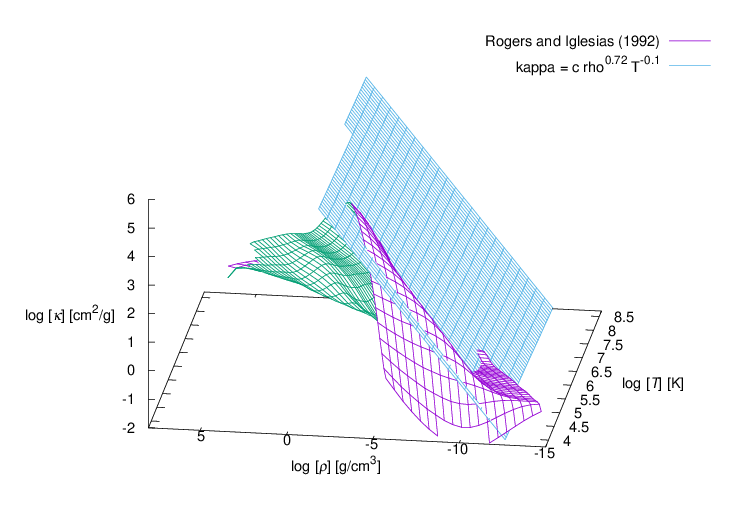}}
    \caption{Opacity approximation}
    \label{fig:Rho_opacity_approx_}
\end{subfigure}
\hfill
    \begin{subfigure}{0.46\hsize}
    \resizebox{1.0\hsize}{!}{\includegraphics{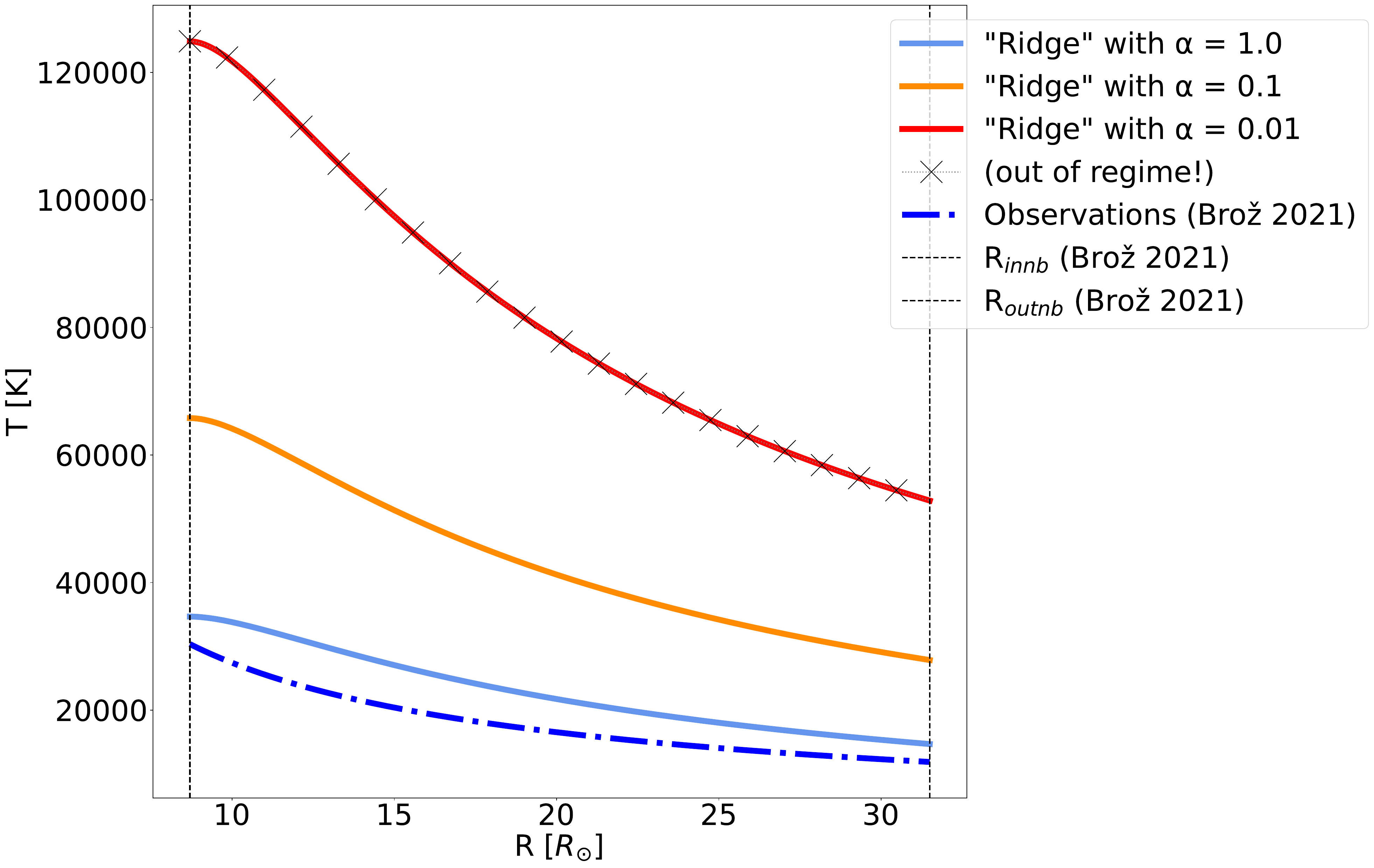}}
    \caption{Temperature profile}
    \label{fig:Pgas_Rho_T}
\end{subfigure}
\hfill
\caption{Same as Fig.~\ref{figs:Pgas_Krammer}, assuming "Ridge" opacity.}
\label{figs:Pgas_Rho}
\end{figure*}

We optimised an approximation of the ridge (of the 2D opacity function of \citet{1992ApJ...401..361R}), because it corresponds to the density and temperature ranges inferred from the \citep{2021A&A...645A..51B} (Fig.~\ref{fig:Rho_opacity_approx_}). We note that, due to the quickly changing slope of the 2D opacity function, the final values of the parameters strongly depend on the choice of the intervals (of temperature and density): 
\begin{equation} \label{eq:ridge_op}
\kappa = 10^{7.67} \rho^{0.72} T^{-0.1}\,.
\end{equation}
The results are shown in Fig.~\ref{figs:Pgas_Rho}.

The "Ridge" model generates profiles consistent with the assumptions of the model down to a low order of magnitude of $\alpha$. Below approximately $\alpha \leq 0.01$ (as seen in Fig.~\ref{fig:Pgas_Rho_T}), the model implies temperatures in the disc beyond the interval, where the "Ridge" opacity can be considered valid, hence we must reject the model for those values of $\alpha$. We note that we tested even lower orders of magnitude and the tendency held.

The temperature, surface density and pressure scale height profiles show even higher values than in the case of \kvthrd{Kramers} opacity (and so also the calculated pressure profiles). The power-law dependence in the outer part of the temperature is significantly worse than in other inspected models\kvthrd{($T \propto r^{-1.08}$, Tab.~\ref{tab:gasdommodel})}. 
\kvthrd{A}s in the case of the \kvthrd{Kramers} model, the observed aspect ratio can be reached in a hydrostatic equilibrium.

\subsubsection{"High-temperatures" opacity}

\begin{figure}[h]
    \centering
    \resizebox{0.47\hsize}{!}{\includegraphics{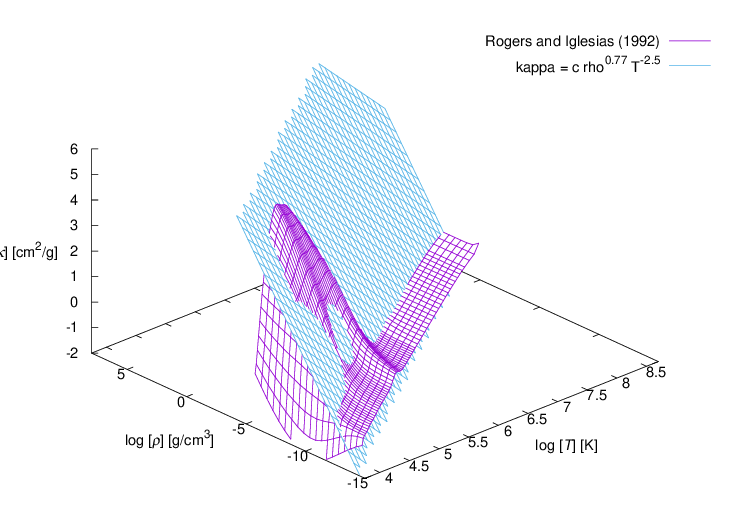}}
    \caption{\kvsec{Same as Fig.~\ref{fig:rogandigl and krammer} but} suitable for regions with high temperatures.}
    \label{fig:High temp_approx+opacity}
\end{figure}

The other tested opacities all tend to have significantly higher midplane temperatures than the observed profiles, hence we decided to find an approximation to the region of higher temperatures (Eq.~(\ref{eq:approx_hightemp}) and Fig.~\ref{fig:High temp_approx+opacity}):
\begin{equation} \label{eq:approx_hightemp}
\kappa = 10^{18.6} \rho^{0.77} T^{-2.5}\,.
\end{equation}
The models are consistent with the assumptions under which they were derived for all tested values of $\alpha$. The profiles generated by assuming this opacity prescription are quantitatively similar to the \kvthrd{Kramers} model, \kvsec{actually considering the similarity in results, of the prescriptions and the proximity in parameter space we should consider it two approximations of the same opacity regime (the detailed values of the parameters are also a consequence of the convergence.)}
\subsection{Models where $P_{\rm g} \approx P_{\rm r}$ applied to the $\beta$~Lyr~A disc}

Here, we apply profiles from Eqs. (\ref{mod_shak_sun_approx_start})-(\ref{mod_shak_sun_approx_end}).

\subsubsection{\kvthrd{Kramers} and "Ridge" opacities}

In this class of models we first experimented with two of the opacity approximations applied in the gas pressure dominated class (in Tab.~\ref{tab:gasdommodel} designated as \kvthrd{Kramers} and "Ridge").

\begin{figure*}[h]
    \centering
    \begin{subfigure}{0.47\hsize}
    \resizebox{\hsize}{!}{\includegraphics{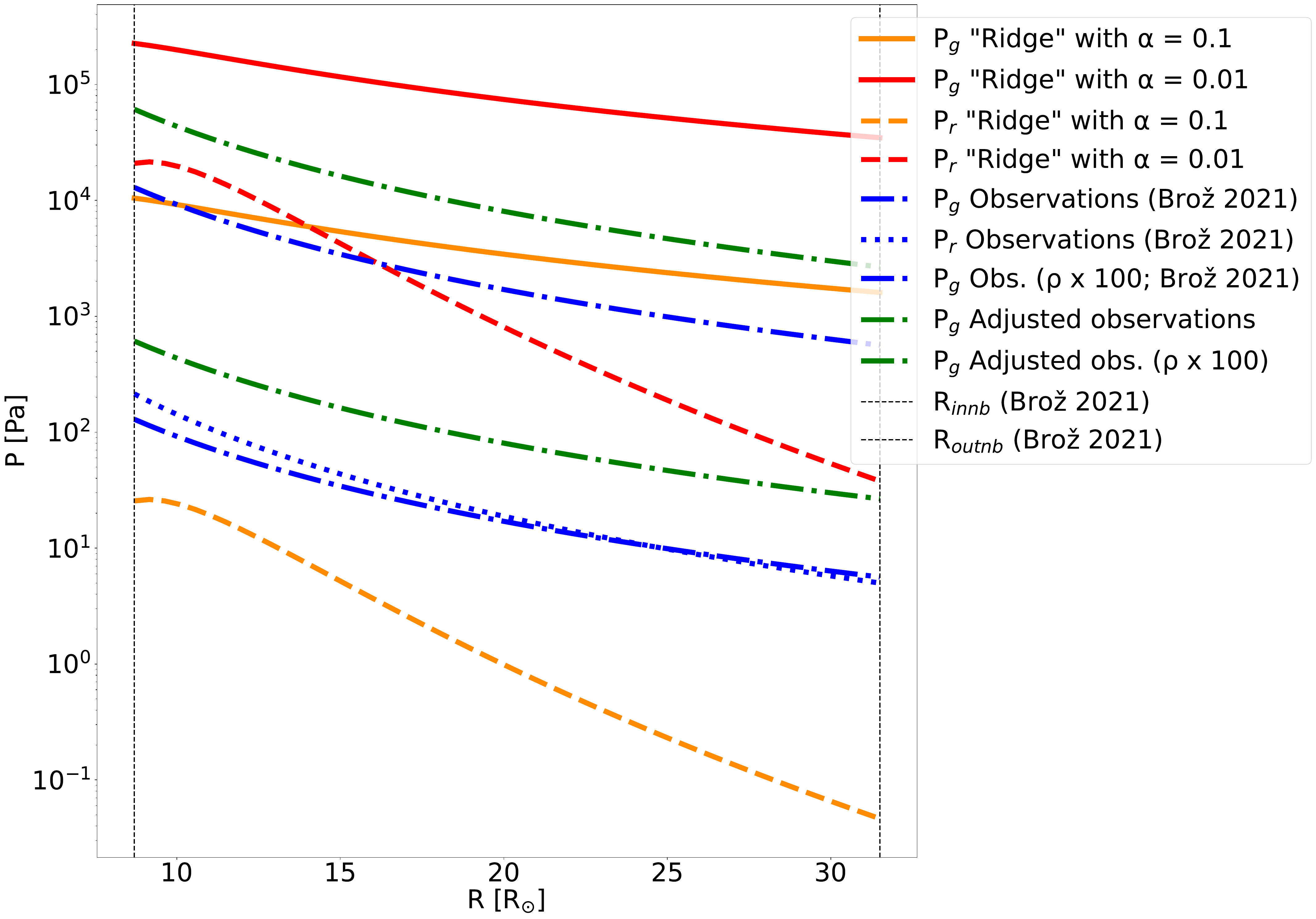}}
    \caption{Pressure profile}
    \label{fig:Pap_Rho_P}
\end{subfigure}
\hfill
    \begin{subfigure}{0.47\hsize}
    \resizebox{\hsize}{!}{\includegraphics{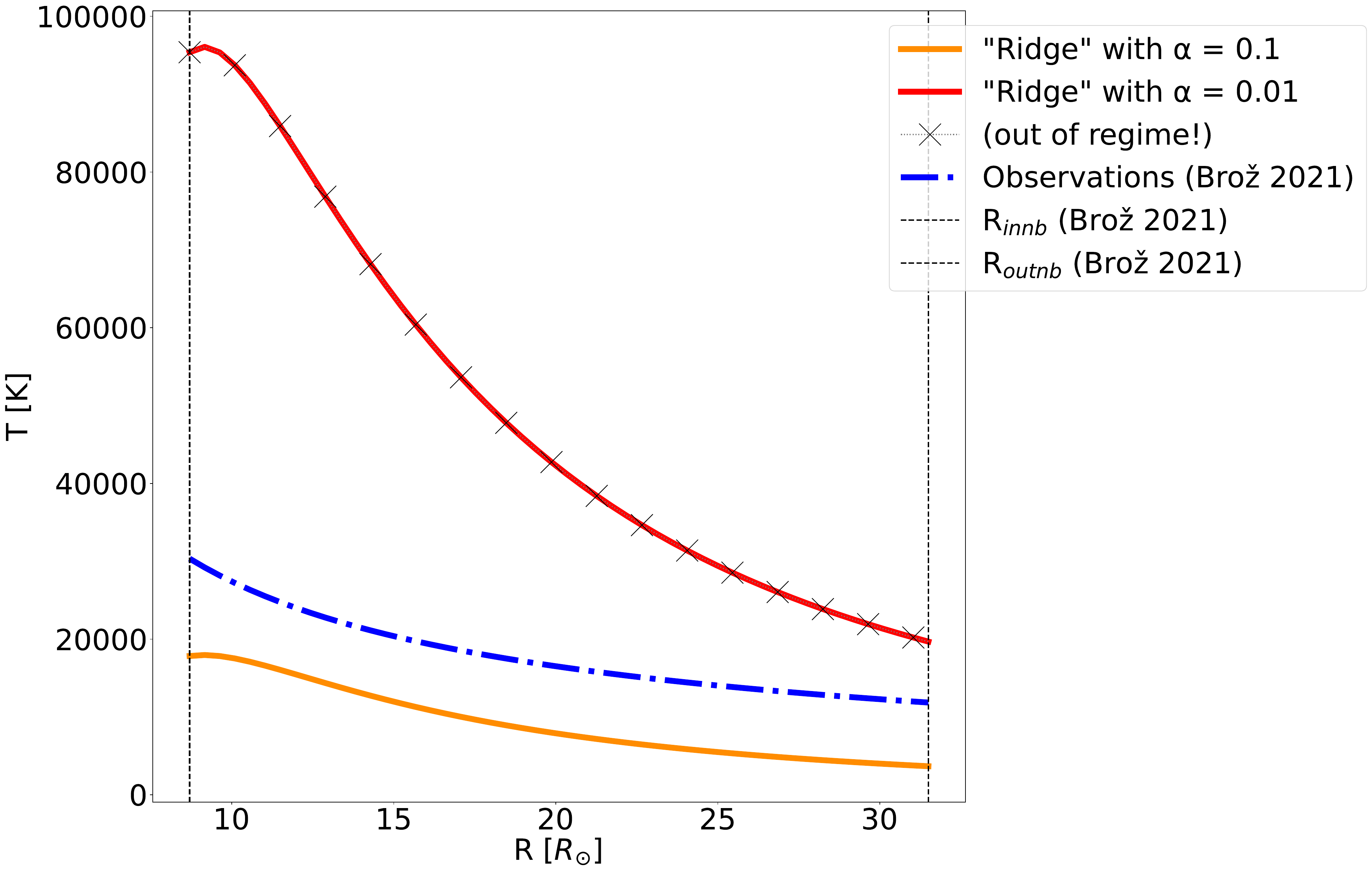}}
    \caption{Temperature profile}
    \label{fig:PaP_Rho_T}
\end{subfigure}
\hfill
\caption{Same as Fig.~\ref{figs:Pgas_Krammer}, assuming $P_{\rm g} \approx P_{\rm r}$ and "Ridge" opacity.}
\label{figs:PaP_inconsistency}
\end{figure*}

Radial profiles of hydrodynamical quantities of both models are inconsistent with the assumptions under which they were derived. Fig. \ref{figs:PaP_inconsistency} demonstrates this on the case of the "Ridge" opacity. For $\alpha \geq 0.01$ there is an inconsistency in pressure profiles, the class defining assumption $P_{g \rm} \approx P_{r \rm}$ is not fulfilled. The gradual changes in the pressure profiles for different values of $\alpha$ in Fig.~\ref{fig:Pap_Rho_P} could indicate that for some even lower $\alpha$  the pressure profiles might become consistent, but the temperature profiles in Fig.~\ref{fig:PaP_Rho_T} show for low $\alpha$($\leq 0.01$) temperatures much higher than the interval for which the opacity approximation may be considered valid. We reject both models completely.

\subsubsection{"Inverse problem" model}

\begin{figure}[h]
    \centering
    \resizebox{0.49\hsize}{!}{\includegraphics{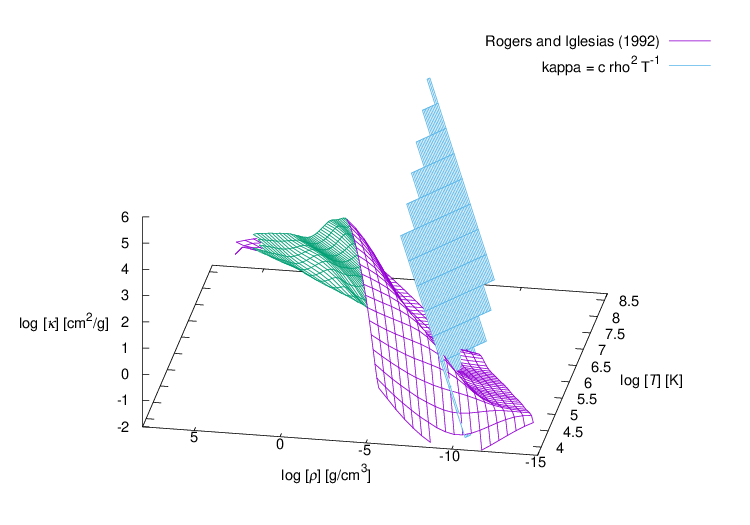}}
    \caption{
    Same as \kvsec{Fig.~\ref{fig:rogandigl and krammer}},
    but where the parameters $\kappa_0$, $A$, $B$ were chosen so that
    the profiles from the modified Shakura-Sunyaev model, assuming $P_{\rm g} \approx P_{\rm r}$,
    are consistent
    (i.e., without being constrained by \citealt{1992ApJ...401..361R}).}
    \label{fig:inverse_approx+opacity}
\end{figure}

One can ask the inverse question; which opacity is necessary for $P_{g \rm} = P_{r \rm}$? We derived formulas for $P_{g \rm}$ and $P_{r \rm}$ with general constants $A$ and $B$. Demanding that exponents of the same variable/parameter match lead to a set of four linear equations. Only two of them are linearly independent, hence a unique solution cannot be obtained. So we searched for an opacity approximation that would ensure consistency with the  $P_{\rm g} \approx P_{\rm r}$ at least in a part of the disc and reached the approximation shown in Fig.~\ref{fig:inverse_approx+opacity}.
\begin{equation} \label{eq:inv_problem eq.}
    \kappa = 6 \cdot 10^{23} \rho^{2} T^{-1}\,.
\end{equation}
We note that the approximation of the opacity is rather 'aggressively' adjusted to achieve the goal and does not correspond to the 2D opacity function very well.

The obtained model is consistent for $\alpha = 0.1$ in the inner part of the disc (up to approximately $15 R_{\odot}$). As in the case of the consistent gas pressure dominated models the generated densities \kvsec{($ < 10 \times$)}, temperatures \kvsec{($ \ge 2 \times$)} and pressures \kvsec{($ \approx 100 \times$)} are significantly higher than in the observed profiles, as well as the aspect ratio can be reached in a hydrostatic equilibrium. A careful consideration of the results indicates (for example the jump in temperatures when going from $\alpha = 0.1$ to $0.01$) that the interval for which a partial consistency is possible is rather narrow and the consistency is "unstable".

\subsection{Radiation pressure dominated models of the $\beta$~Lyr~A disc}

Here we apply profiles from Eqs.~(\ref{mod_shaksun_rad_start})-(\ref{mod_shaksun_rad_end}).

\subsubsection{\kvthrd{Kramers} and "Ridge" opacities}

Again, we first experimented with implementing the same opacity approximations that were used in the other classes (those designated as \kvthrd{Kramers} and "Ridge" in Tab.~\ref{tab:gasdommodel}). Similarly to the previous class ($P_{g \rm} \approx P_{r \rm}$), both models must be rejected on the basis of inconsistencies between generated profiles and assumptions used in the derivation, the "Ridge" model pressure profiles are in contradiction to the class defining $P_{g \rm} \ll P_{r \rm}$ assumption for $\alpha \geq 0.01$, the differences between profiles for different values of $\alpha$ indicates that for even lower values of $\alpha$ the assumption might be satisfied, but the temperature profiles are too high for the opacity regime and seem to increase as $\alpha$ decreases (hence the inconsistencies "grow" in opposite directions). In \kvthrd{Kramers} model pressure profiles are contradictory almost irrelevant of $\alpha$, the temperature profiles are on the fringe of the opacity regime.

\subsubsection{"Extreme temperatures" model}

\begin{figure}[h]
    \centering
    \resizebox{0.49\hsize}{!}{\includegraphics{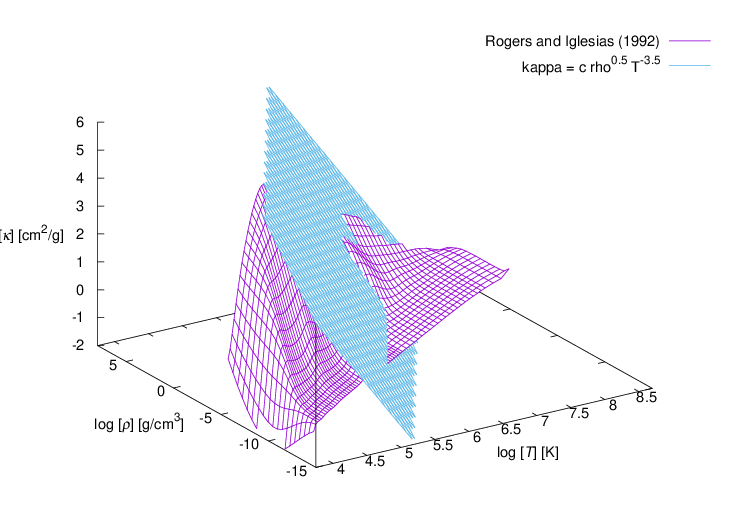}}
    \caption{Same as Fig.~\ref{fig:rogandigl and krammer}, for "Extreme temperatures" opacity.}
    \label{fig:Extremetemp_approx+opacity}
\end{figure}

Following the intuition that radiation pressure-dominated models should be valid for hotter discs, we searched for an opacity approximation in the $T \approx 10^{5},10^{5.5}$ K region (black hole accretion discs have temperatures of the order of $10^6$K so we wanted to stay below that). This model is the result of these considerations. The opacity approximation designated as "Extreme temperatures" is plotted in Fig.~\ref{fig:Extremetemp_approx+opacity}. 
\begin{equation}
   \kappa = 6 \cdot 10^{23} \rho^{0.5} T^{-3.5} \, .
\end{equation}
The $P_{\rm g} \ll P_{\rm r}$ condition is satisfied for $\alpha$ close to 1.0, and the temperatures are reasonable for the regime. Again the intervals of input parameters that lead to consistent solutions are rather narrow and the consistency cannot be considered robust (\kvsec{it is "un}stable").

\begin{figure*}[h]
    \centering
    \begin{subfigure}{0.49\hsize}
    \resizebox{1.0\hsize}{!}{\includegraphics{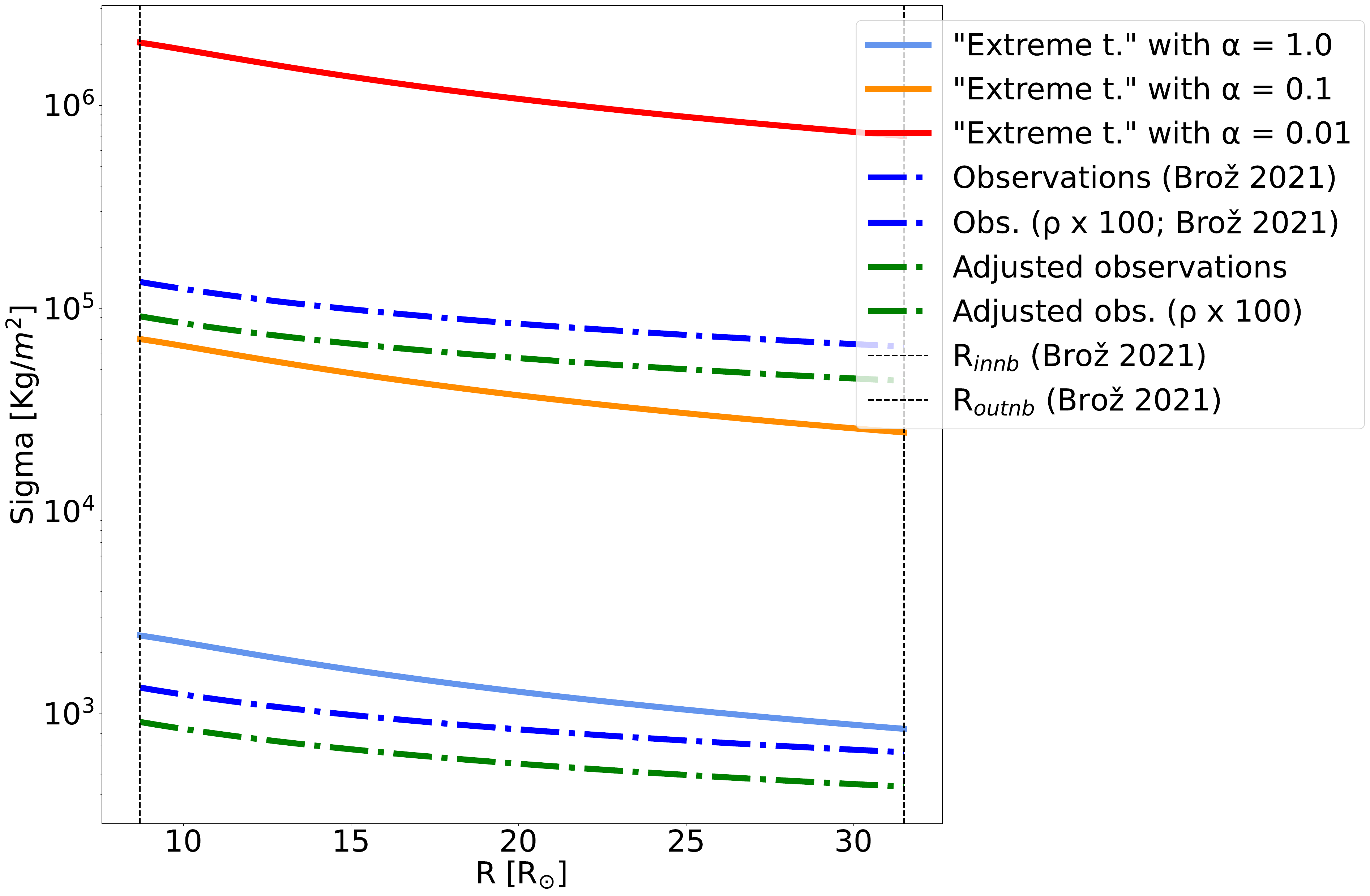}}
    \caption{Surface density $\Sigma$}
    \label{fig:Prad_extrm_Sigma}
\end{subfigure}
\hfill
    \begin{subfigure}{0.49\hsize}
    \resizebox{1.0\hsize}{!}{\includegraphics{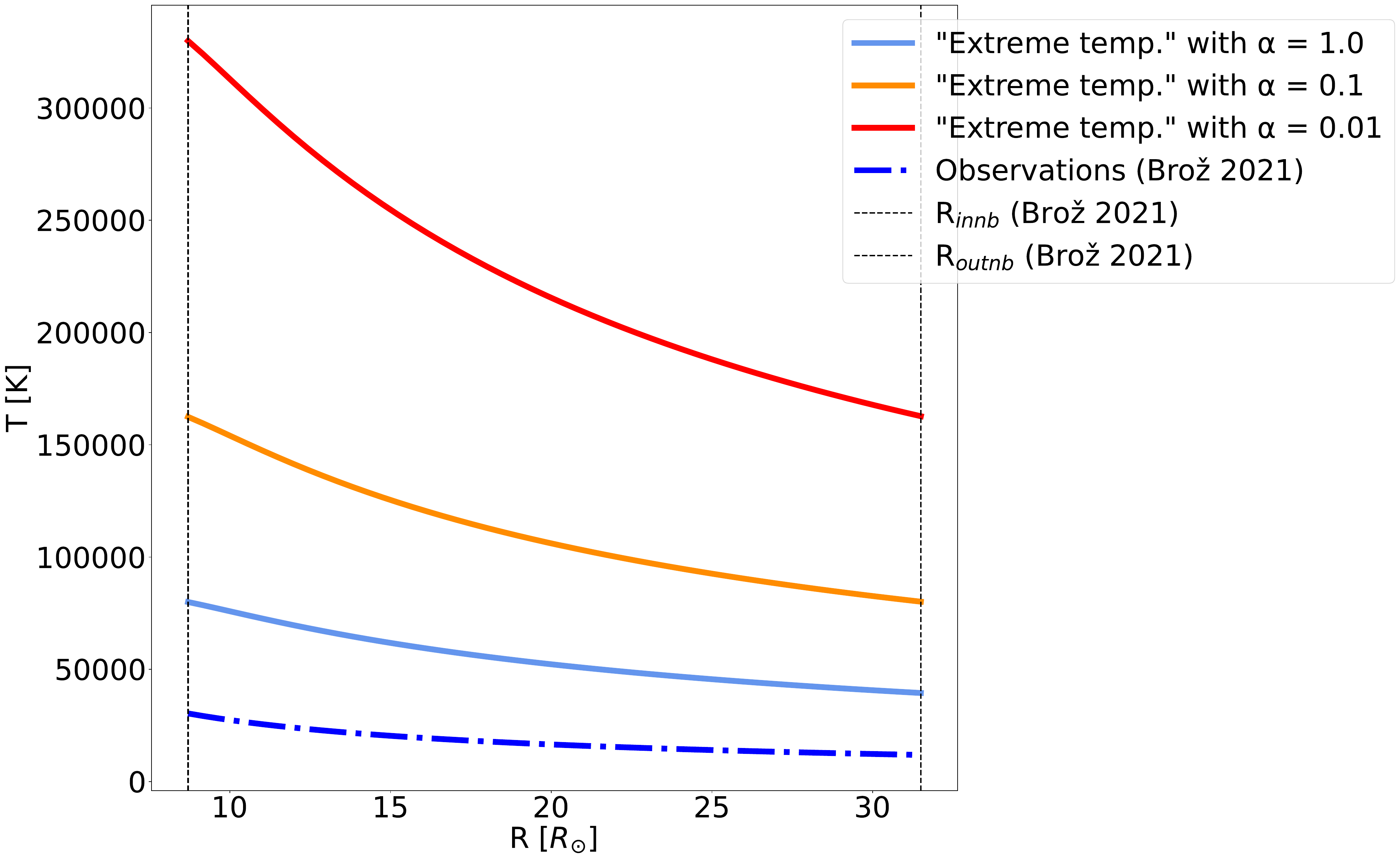}}
    \caption{Temperature $T$}
    \label{fig:Prad_extrm_T}
\end{subfigure}
\hfill
\begin{subfigure}{0.49\hsize}
    \resizebox{\hsize}{!}{\includegraphics{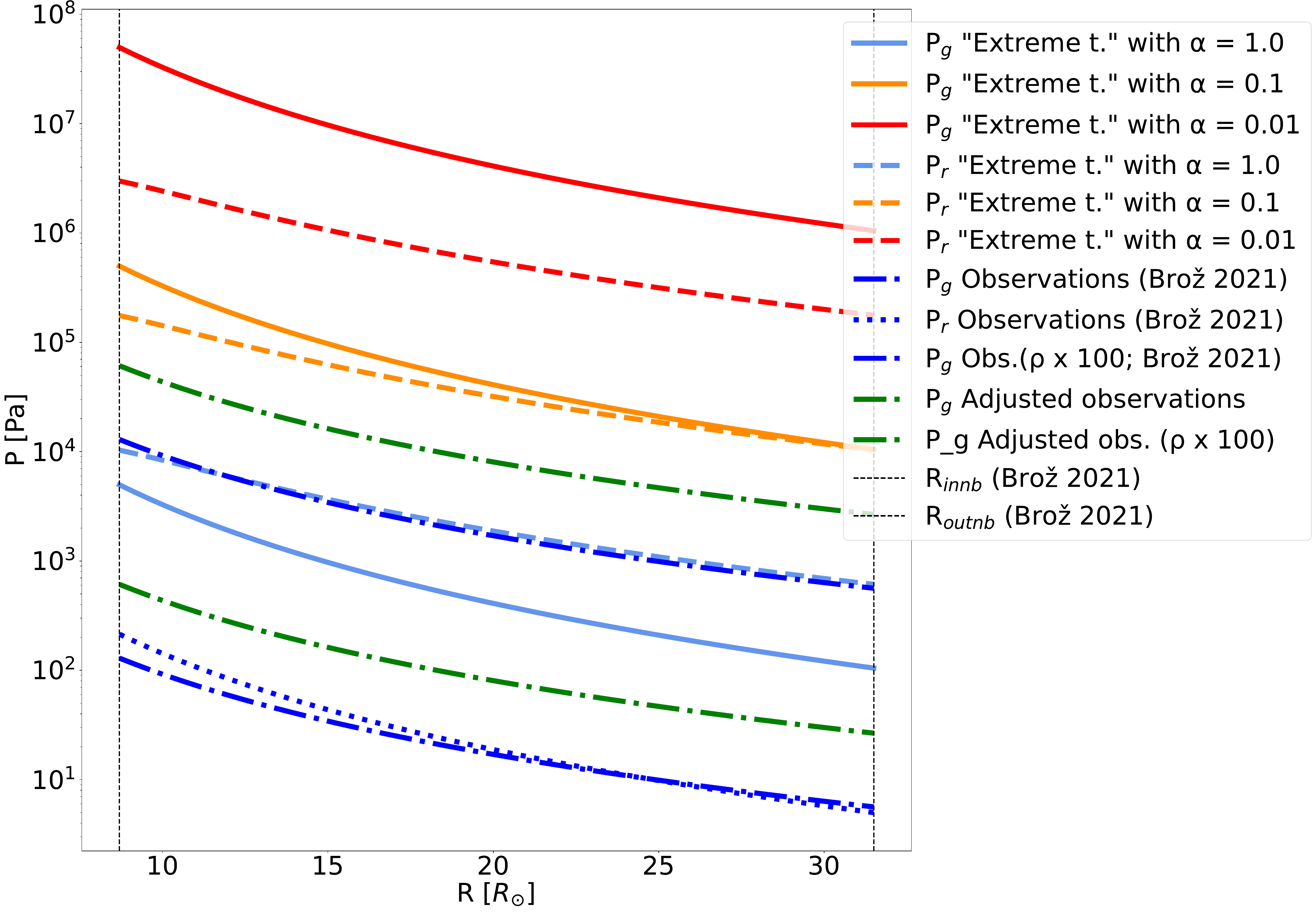}}
    \caption{Gas pressure $P_{g \rm}$ and radiation pressure $P_{r \rm}$}
    \label{fig:Prad)_extrm_P}
\end{subfigure}
\hfill
\begin{subfigure}{0.49\hsize}
    \resizebox{\hsize}{!}{\includegraphics{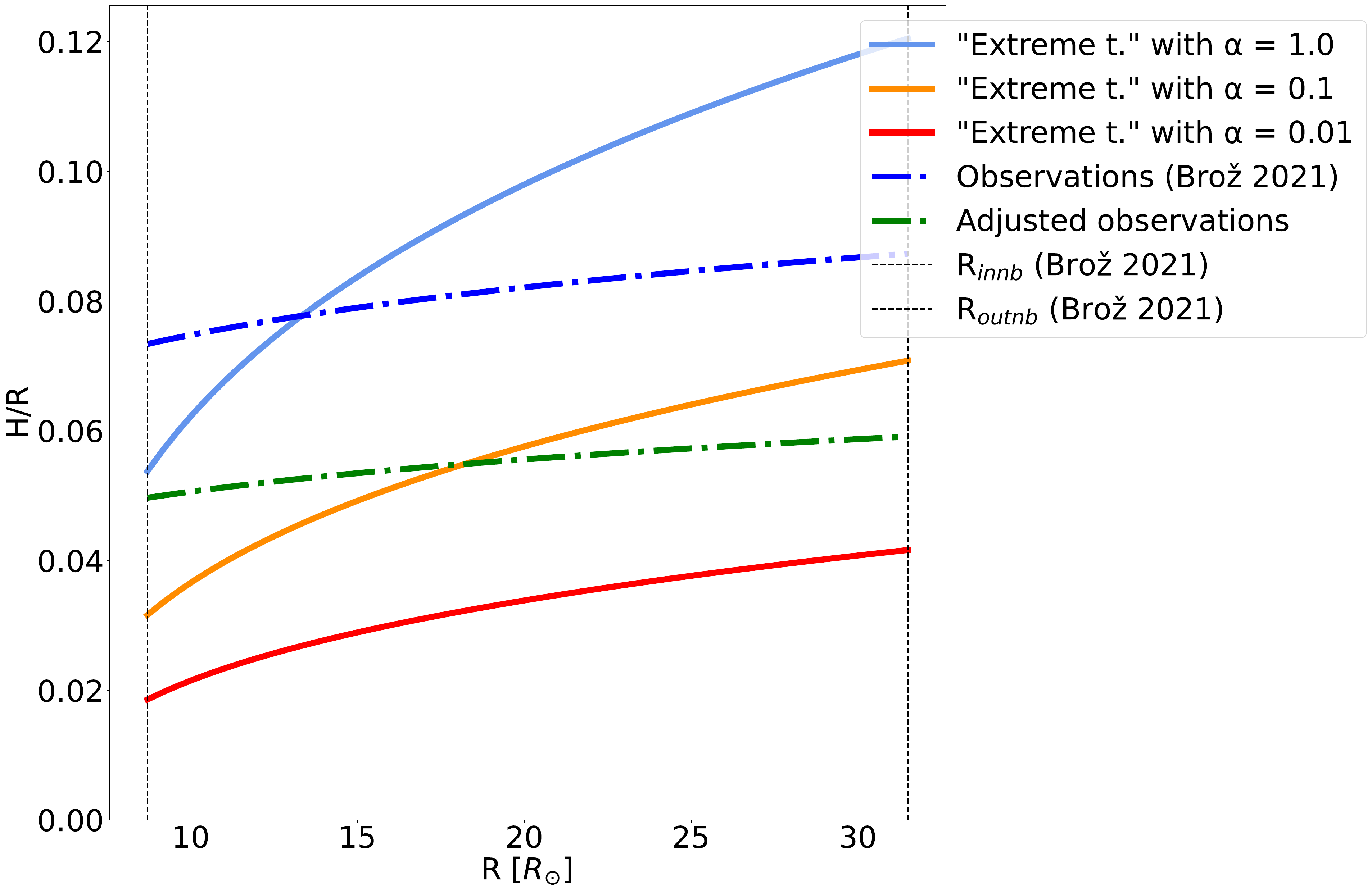}}
    \caption{Aspect ratio $H/r$}
    \label{fig:Prad_extrm_Aspect}
\end{subfigure}
\caption{Same as Fig.~\ref{figs:Pgas_Krammer}, assuming $P_{\rm g} \ll P_{\rm r}$ and "Extreme temperatures" opacity.}
\label{figs:Prad_extrm}
\end{figure*}

The predictions of the consistent $\Sigma$, $T$ and $H$ profiles (plotted in Fig. \ref{figs:Prad_extrm}) indicate a possibility to construct a disc that is somewhat hotter than predicted by other models and less dense (but still more than the observed profiles) but reaches comparable pressure scale heights.
A counter-intuitive result from this model is that higher temperatures (lower $\alpha$) seem to produce a thinner disc. Here we emphasise that this aspect of the model should not necessarily be taken to be real, since it is for cases that are \kvsec{inconsistent} with the $P_{\rm g} \ll P_{\rm r}$ assumption and around $\alpha \approx 1.0$, where the assumption is met, the effect is very weak; $T \propto \alpha^{0.3}$.

\section{Alternative solutions} \label{sec:alter_sol}

\kvthrd{In Sec.~\ref{sec:52}, \mb{the temperature problem was introduced and we} discussed what \mb{is} our \mb{preferred} solution\mb{. Here,} alternative \mb{solutions} are \mb{tested}.}

\subsection{\kvthrd{Solution 2: Opacity table of \citet{1994ApJ...427..987B}}}

\kvthrd{We \mb{tested} the uncertainty that is implied by the specific choice of opacity table. We recomputed the same simulation but now using the opacity table by \citet{1994ApJ...427..987B}. The opacity profile remained qualitatively the same (notably\mb{,} it did not contain any transitions), but \mb{it was} up to ten times higher. \mb{Nevertheless,} the influence on the temperature profile was less than 10\%.}

\subsection{\mb{Solution 3: Irradiation}}

\kvthrd{Considering the limited extent of the disc, it was to be checked that we weren't crucially underestimating the influence of irradiation from the central star. In particular, the model does not include any edge-on irradiation. We \mb{thus} artificially increased the temperature of the central star to $T_{\star} = 50000\,\text{K}$ and set the albedo to zero ($A = 0$). The effect on the temperature profile was negligible.}

\subsection{Solution 3: Vertical opacity profile} \label{vert_op_prof}

The $C_{\rm k}$ parameter (from Eq.~(\ref{eq:tau_opt}) affects the vertical opacity profile.
This parameter is typically set to 0.6 to account for a drop in opacity above the midplane
in cool discs \citep{2017A&A...606A.114C}.
\kvsec{The disc studied in this work may not be considered cool and hence the value of this parameter remains an open question.}
In principle, a single opacity law may govern the whole vertical opacity profile ($C_{\rm k} = 1$).
If an opacity law is inversely proportional to temperature,
then in a disc where the temperature decreases with distance from the midplane,
the opacity can theoretically grow.
We tested the latter possibility by setting $C_{\rm k} = 1.5$.

\begin{figure}
    \centering
    \includegraphics[width=9cm]{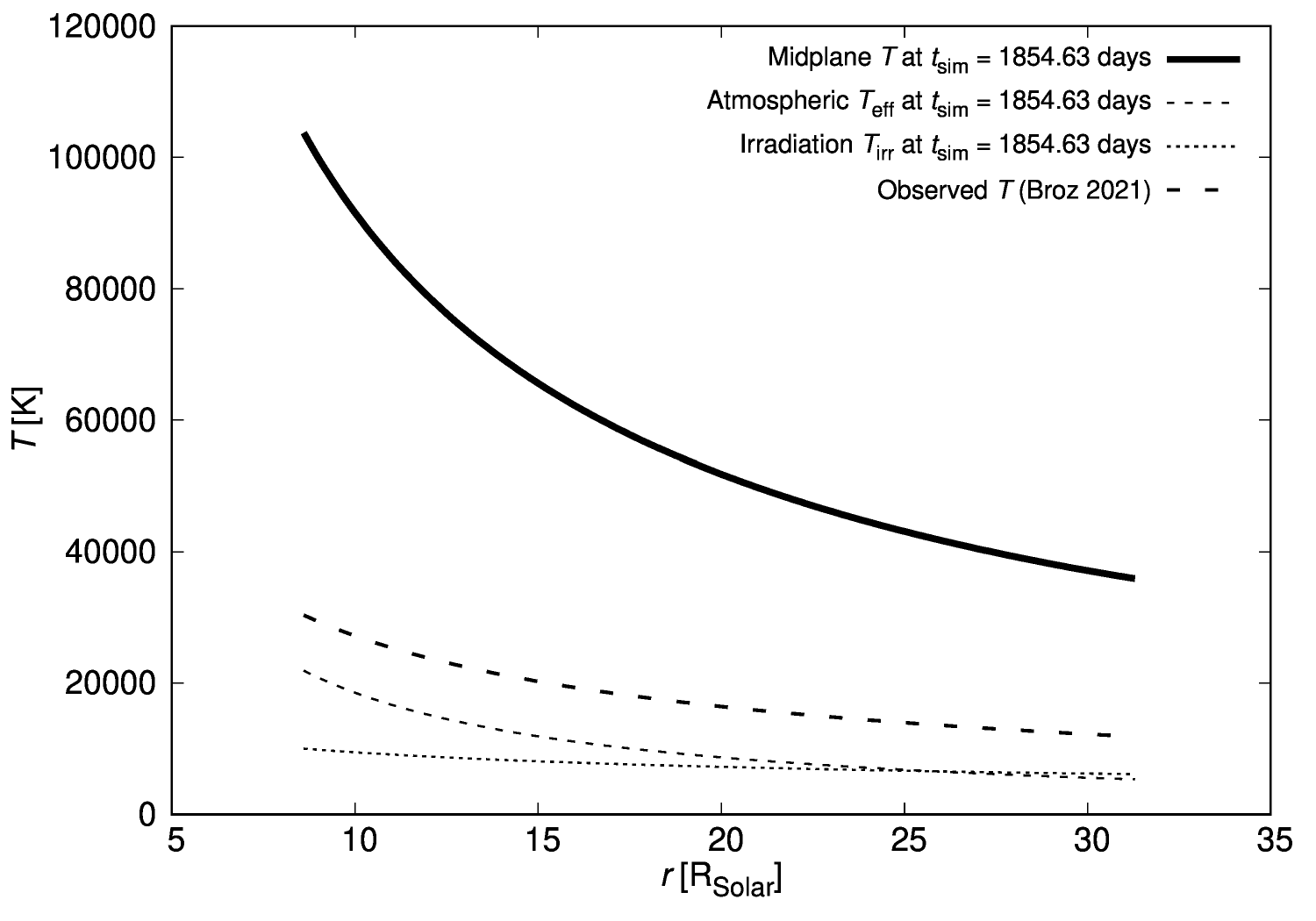}
    \caption{
    Same as Fig.~\ref{fig:zhu_gen} (top), but with $C_{\rm k} = 1.5$ (opacity increases above the midplane) and $A = 0$.
    The increased isolation causes an increase in midplane temperatures. 
    $T_{\rm eff}$ and  $T_{\rm irr}$ coincide well in the outer part of the disc.
    }
    \label{fig:gastem_ck}
\end{figure}

The temperature profiles plotted in Fig.~\ref{fig:gastem_ck} show a substantial increase in midplane temperatures
and a small increase in atmospheric temperatures (over 100000\,K at the inner boundary and just under 40000\,K at the outer boundary).
The opacity increase in the vertical direction acts as an extra layer of isolation,
that is the optical depth of the midplane increases. 
$T_{\rm eff}$ and $T_{\rm irr}$ coincide well in the outer part of the disc,
but both are lower than the observations.
In the inner part of the disc, $T_{\rm eff}$ is even higher than $T_{\rm irr}$.
This would break our assumption that the irradiation temperature serves as the thermodynamic limit to the atmospheric temperature.

The increase in temperature causes a thicker disc. The aspect ratio is thus in very good agreement with the observations.

The increase in temperature also implies an increase in viscosity
(in Eq.~\ref{eq:kin_visc_alpha_param}\kvsec{ $H \propto \sqrt{T}$ and $c_{\rm s} \propto \sqrt{T}$}).
The higher the viscosity, the more effective the transport of angular momentum and greater radial velocities;
at the inner boundary velocities reach about $700\,{\rm m}\,{\rm s}^{-1}$.
A faster transport of matter is coupled with a decrease in surface densities (by $\approx 10 \%$).

\subsection{Solution 4: High $\alpha$ viscosity parameter} \label{high_alpha}

Using the analytical models, we demonstrated that a higher viscosity parameter $\alpha$ implies lower temperature profiles.
Here, we present results for the limit case of $\alpha = 1.0$ and also $\alpha = 1.5$ (which is beyond the classical limit).

\citet{1973A&A....24..337S} introduced the parametrisation and reasons for the limit $\alpha \leq 1.0$, but the argumentation is based on a single source of viscosity. In fact, without a full theory of viscosity on the microscopic scale, this limit is not guaranteed. It is to be considered that several mechanisms of the angular momentum transfer may be present at once. 

Even macroscopic phenomena may effectively add to the value of $\alpha$, e.g., spiral waves induced by tidal forces propagating from the outer boundary inwards will travel at roughly the sound speed, which is slower than the orbital motion in the disc. This differential rotation leads to an angular momentum loss \citep{SpiralShocksTheo}. This is possible when the 3:1 Lindblad resonance is located within the radial extent of the disc. According to \citet{SpiralShocksTheo}, this can be true for discs of $q < 0.33$. The $\beta$~Lyr~A disc ($q = 0.223$; \citealt{2021A&A...645A..51B}) is within the limit. The 1D simulations cannot resolve higher dimensions, but we found a reasonably good correspondence between the sound-speed timescale of adjusted observations and the binary period at the outer disc boundary. We take this as an additional argument that spiral waves may be present.

For both values of $\alpha$, models stopped showing temporal evolution within a year and a half of the simulation time in all studied quantities.

High $\alpha$ means an effective angular momentum transport, combined with the fixed mass transfer implied an order of a magnitude lower surface densities (but still almost 10 times the observed profile) coupled with high radial velocities (up to $5000\,{\rm m}\,{\rm s}^{-1}$).

The 10 (resp. 15) times increase in $\alpha$ resulted in a decrease by almost a half in the midplane temperatures.
For the gas-pressure-dominated model with \kvthrd{Kramers} opacity,
we derived the exponent over $\alpha$ to be $\alpha^{-\frac{1}{5}}$ and $10^{-\frac{1}{5}}=0.630957$.
The atmospheric temperatures $T_{\rm eff}$ did not change significantly;
this is coupled with the aspect ratio, which decreased to $H/r \approx 0.06$.

\subsection{Solution 5: Radiation pressure}

In our initial considerations of the $\beta$~Lyr~A system, it was not clear how important the radiation pressure is for the dynamics of the system. The analytical models were mostly only consistent for the gas-pressure-dominated class and the numerical model used an ideal gas equation of state. We thus implemented a simple modification to the FARGO\_THORIN code by \citet{2017A&A...606A.114C}
to approximately account for additional pressure contribution due to radiation. 

We expect that radiation pressure plays a role mainly in mechanical processes, hence we introduced two new parameters.
\begin{itemize}
    \item The definition of $H$ holds for an ideal gas, but \citet{2021ApJ...910...31M} showed that when the radiation pressure is accounted for, the pressure scale height changes by a specific factor. They also introduced a numerical scheme to implicitly find this parameter. We introduced this factor into our model as a fixed parameter.
    \begin{equation} \label{eq:h_factor}
        H_{\rm r+p} = h H \, ,
    \end{equation}
    where $h$ denotes the dimensionless factor,
    $H$ is the pressure scale height defined for an ideal gas,
    and $H_{\rm r+p}$ is the pressure scale height when radiation pressure is included. 
    \item To approximate the role of the increased pressure in the dynamics, we introduced a second dimensionless factor $p$,
    \begin{equation}
        P_{\rm g} + P_{\rm r} = p P_{\rm g} \,.
    \end{equation}
\end{itemize}

\begin{figure}
    \centering
    \includegraphics[width=9cm]{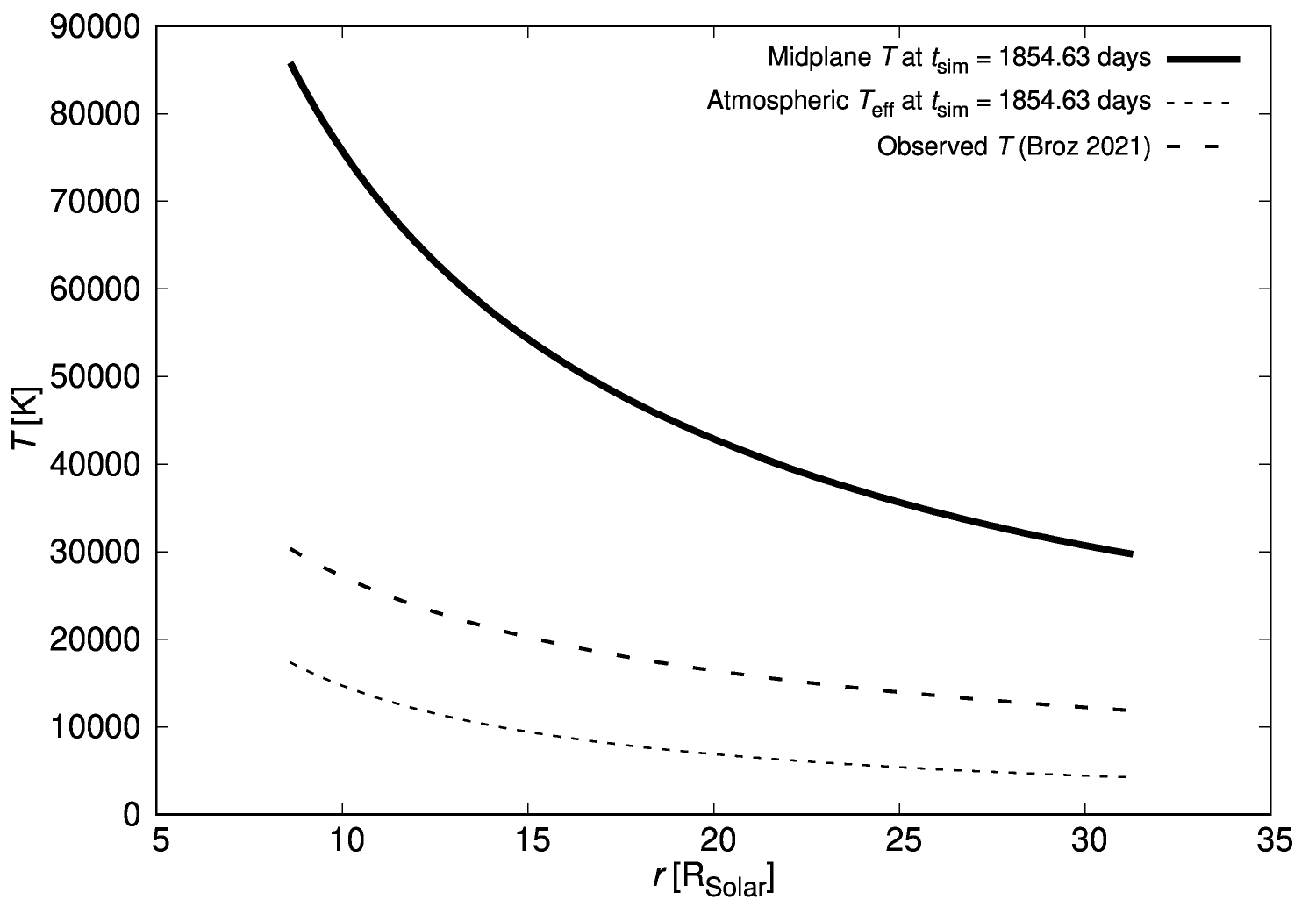}
    \caption{
    Same as Fig.~\ref{fig:zhu_gen}, but for the modified numerical model accounting for the radiation pressure.
    The small addition to pressure causes a decrease in the midplane temperatures, but it does not affect the atmospheric temperatures.
    }
    \label{fig:mod}
\end{figure}

We included these two factors into the fluid equations of the numerical model.
The value of $p = 1.1$ was estimated from pressure profiles generated by the gas-pressure-dominated analytical models
and $h = 1.75$ is the approximate factor between the \mb{observed} aspect ratio and the \mb{adjusted} profile.
The resulting temperature profile is shown in Fig.~\ref{fig:mod}.

We observe coupled phenomena, the aspect ratio increased in a consistent manner with the definition of the $h$ factor (Eq.~(\ref{eq:h_factor})).
The expansion to a higher pressure scale height results in a slight decrease in the midplane temperatures,
which intuitively follows from the ideal gas law.
The atmospheric temperatures remained without significant changes due to a weakening of opacity in the vertical profile
(i.e., lower optical depth, Eq.~(\ref{eq:tau_opt})).
The surface density profile (as well as the radial velocities) remained the same;
considering the increase in $H$, this implies the volumetric density decreased.
The opacity (e.g. \kvthrd{Kramers)} is $\propto \rho$, hence it also decreased. 
The increase in the radial pressure gradient also results in the disc being slightly more sub-Keplerian.
Other studied quantities seem unaffected.

\section{Supplementary figures}

Here, we present supplementary figures of profiles
from our modified Shakura-Sunyaev models described in
Sec.~\ref{sec: results} and Appendix~\ref{sec:other_anal}.

\onecolumn

\begin{figure*}
    \centering
    \begin{subfigure}{0.49\hsize}
    \resizebox{1.0\hsize}{!}{\includegraphics{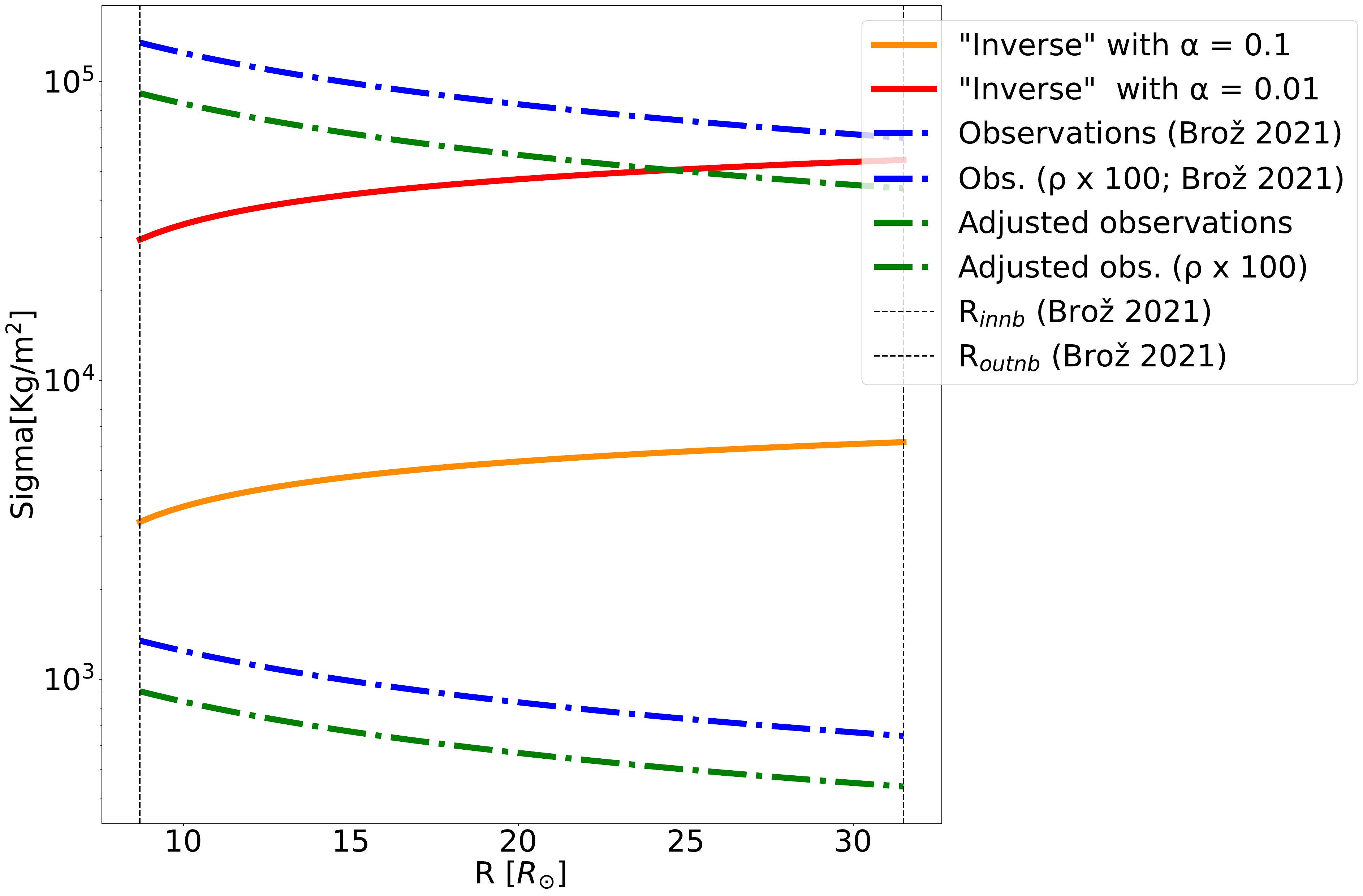}}
    \caption{Surface density $\Sigma$}
    \label{fig:PapP_inverse_Sigma}
\end{subfigure}
\hfill
    \begin{subfigure}{0.49\hsize}
    \resizebox{1.0\hsize}{!}{\includegraphics{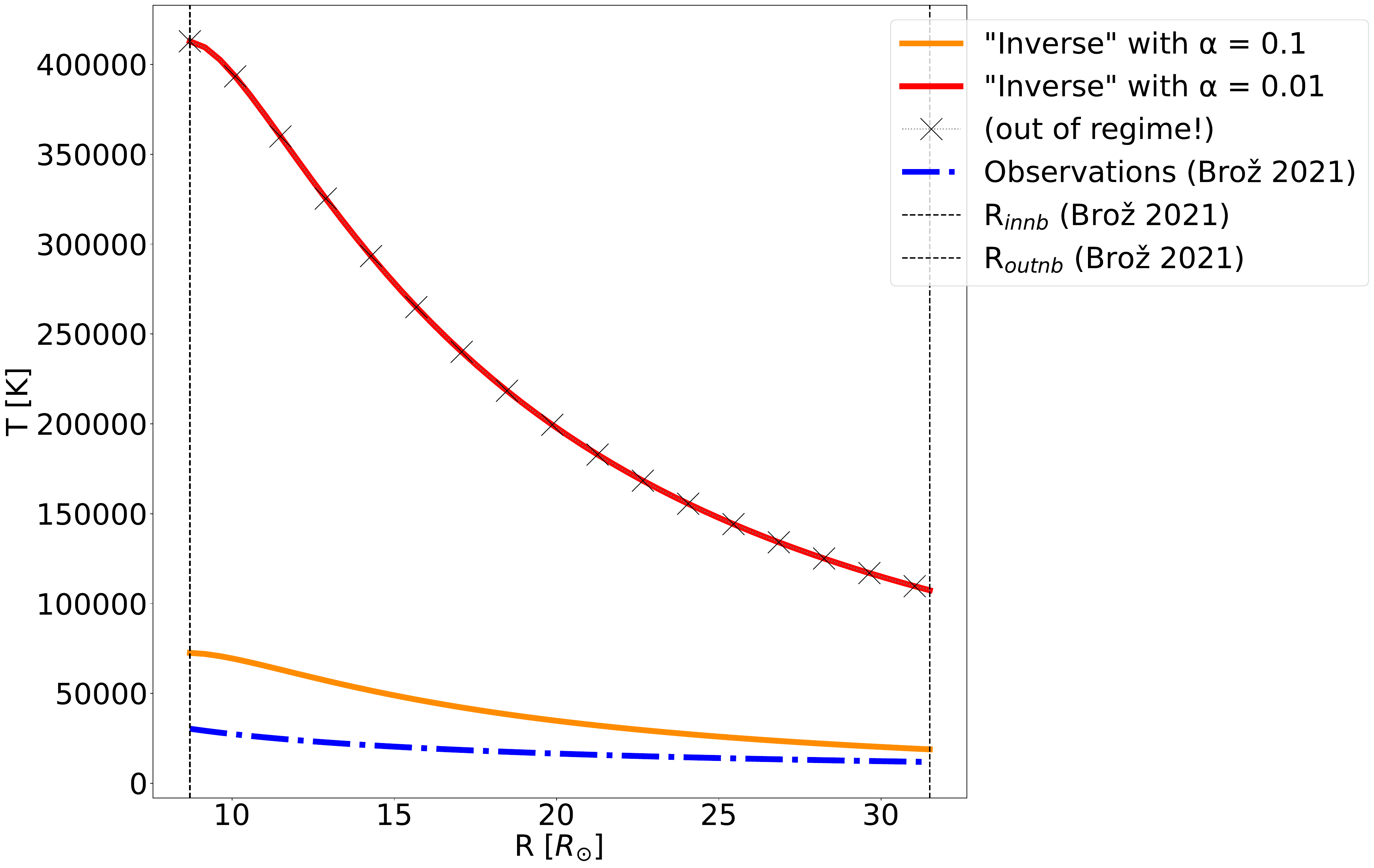}}
    \caption{Temperature $T$}
    \label{fig:PapP_inverse_T}
\end{subfigure}
\hfill
\begin{subfigure}{0.49\hsize}
    \resizebox{\hsize}{!}{\includegraphics{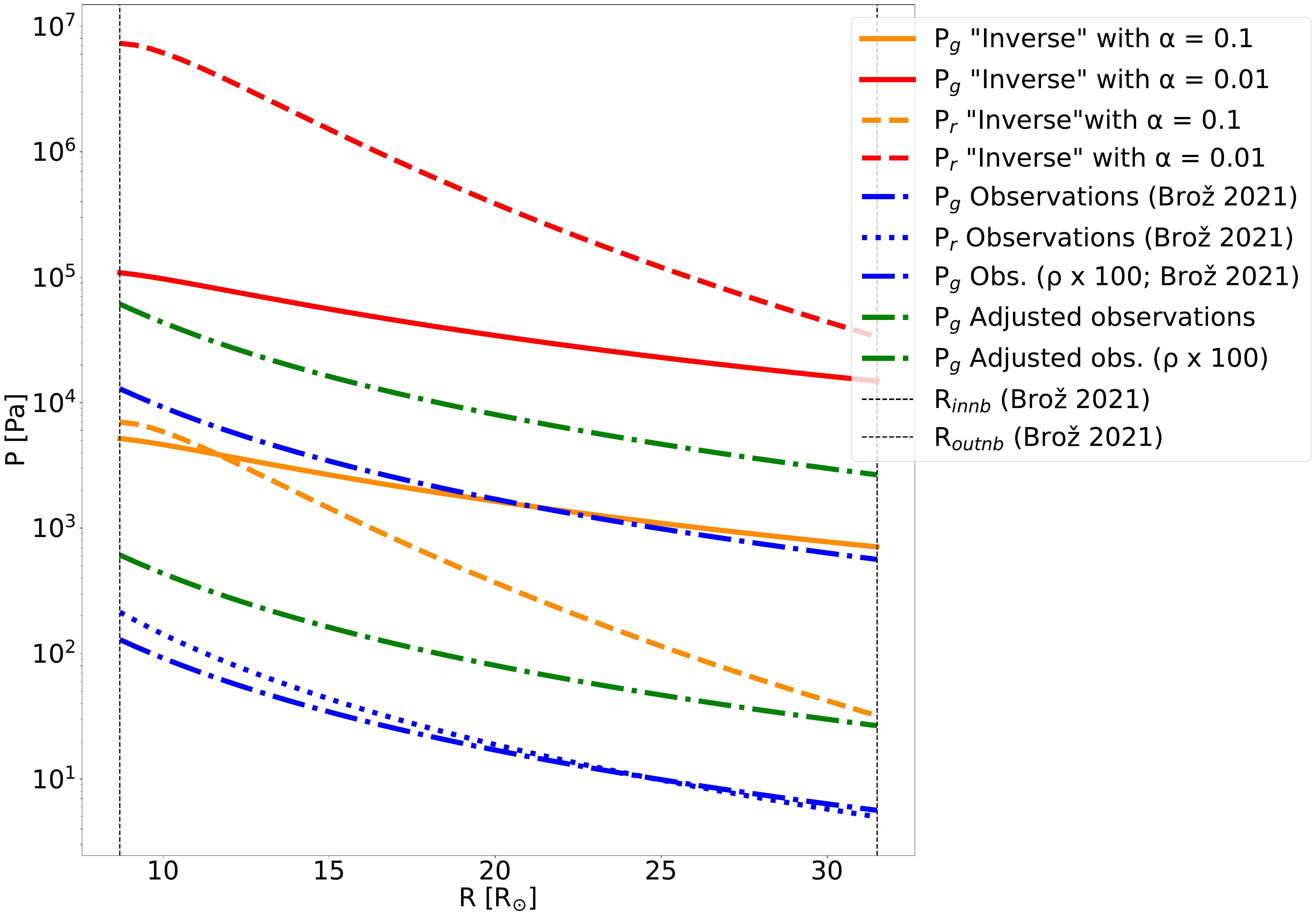}}
    \caption{Gas pressure $P_{g \rm}$ and radiation pressure $P_{r \rm}$}
    \label{fig:PapP_inverse_P}
\end{subfigure}
\hfill
\begin{subfigure}{0.49\hsize}
    \resizebox{\hsize}{!}{\includegraphics{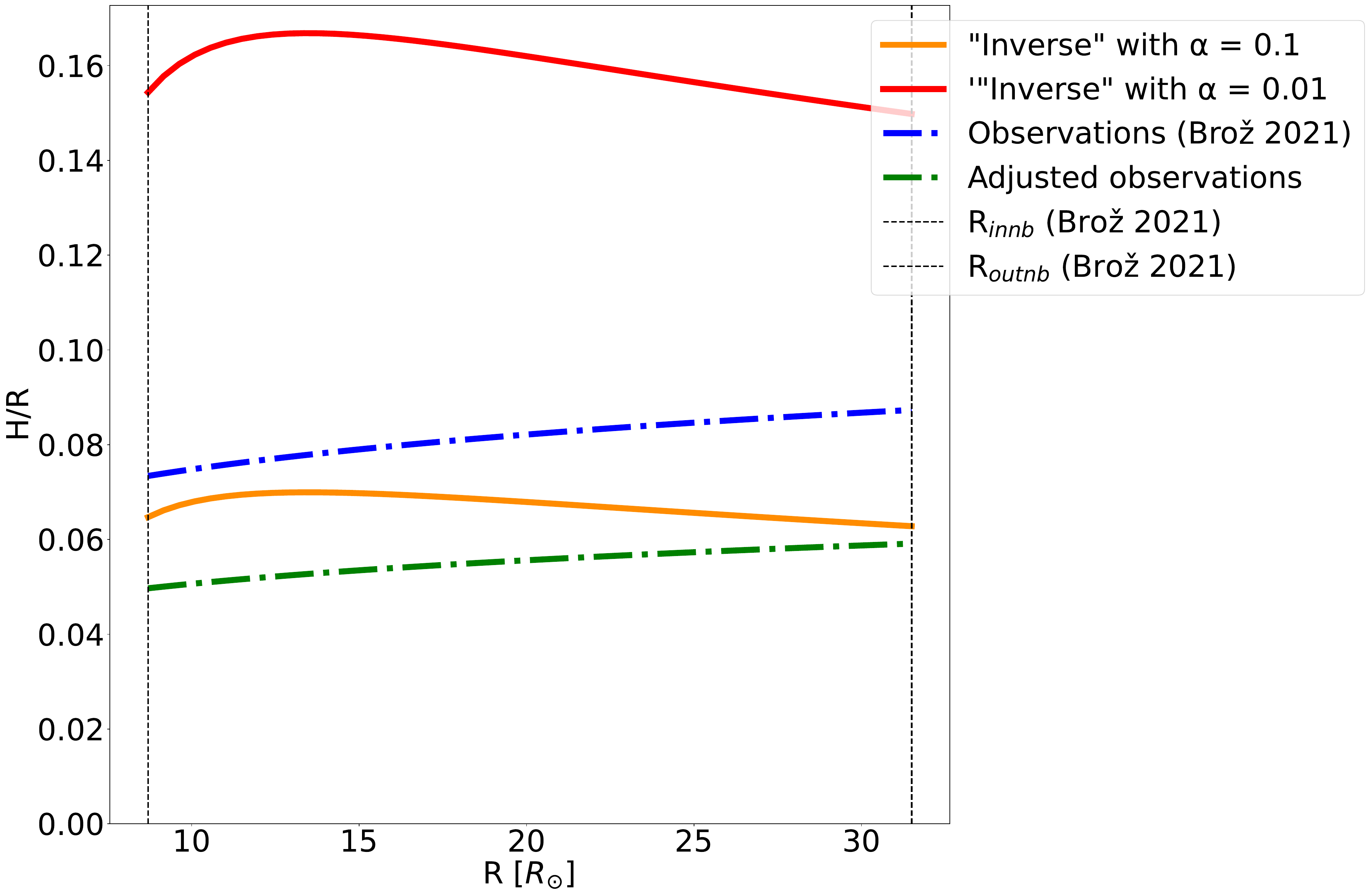}}
    \caption{Aspect ratio $H/r$}
    \label{fig:PapP_inverse_Aspect}
\end{subfigure}
\caption{Same as Fig.~\ref{figs:Pgas_Krammer}, assuming $P_{\rm g} \approx P_{\rm r}$ and "Inverse problem" opacity.}
\label{figs:inverse_problem}
\end{figure*}

\end{appendix}
\end{document}